\documentclass[fleqn]{article}

\usepackage{amsmath,amssymb}
\usepackage{graphicx}

\def\tref#1{(\ref{#1})}
\def\tlabel#1{\label{#1}}

\begin{document}

\title{
{\sf Closed-Form Summation of RG-Accessible
Logarithmic Contributions to Semileptonic B-Decays and Other Perturbative Processes} }

\author{M.R.~Ahmady$^1$, F.A.~Chishtie$^2$, V.~Elias$^3$, A.H.~Fariborz$^4$, \\
N.~ Fattahi$^3$, D.G.C.~McKeon$^3$, T.N.~Sherry$^5$, 
T.G.~Steele$^6$}
\footnotetext[1]{Department of Physics, Mount Allison University, Sackville, NB~~E4L 1E6, Canada}
\footnotetext[2]{
Newman Laboratory of Nuclear Studies, Cornell University, 
Ithaca, NY~~ 14853, USA}
\footnotetext[3]{Department of Applied Mathematics,
The University of Western Ontario,
London, ON~~ N6A 5B7, Canada}
\footnotetext[4]{Dept.\ of Mathematics/Science, State Univ.\ of New York Institute of Technology, Utica, NY~~13504-3050, USA}
\footnotetext[5]{Department of Mathematical Physics, National University of Ireland, Galway, Ireland }
\footnotetext[6]{Department of Physics \& Engineering Physics,
University of Saskatchewan,
Saskatoon, SK~~ S7N 5E2, Canada}
 
\maketitle

\begin{abstract}
For any perturbative series that is known to $k$-subleading orders of perturbation theory, we utilise
the process-appropriate renormalization-group (RG) equation in order to obtain all-orders
summation of series terms proportional to $\alpha^n \log^{n-k}\left(\mu^2\right)$  with $k = \{0,1,2,3\}$, 
corresponding to the
summation to all orders  of the leading and subsequent-three-subleading logarithmic
contributions to the full perturbative series. These methods are applied to the perturbative series
for semileptonic $b$-decays in both $\overline{{\rm MS}}$ and pole-mass schemes, and they result in RG-summed
series for the decay rates which exhibit greatly reduced sensitivity to the renormalization scale $\mu$.
Such summation via RG-methods of all logarithms accessible from known series terms is also
applied to perturbative QCD series for vector- and scalar-current correlation functions, the
perturbative static potential function, the (single-doublet standard-model) Higgs decay amplitude
into two gluons, as well as the Higgs-mediated high-energy cross-section for $W^+W^-\to   ZZ$
scattering. The resulting RG-summed expressions are also found to be much less sensitive to the
renormalization scale  than the original series for these processes.    
\end{abstract}

\section{Introduction}\label{intro_sec}
\renewcommand{\theequation}
{1.\arabic{equation}}
\setcounter{equation}{0}
The renormalization group equation (RGE) has long proven useful as a means of 
improving and extending results obtained from perturbative quantum field theory.  
In addition to giving rise to scale-dependent running parameters (coupling constants 
and masses) and concomitant scale properties ({\it e.g.} asymptotic freedom), the RGE can 
also be utilised to determine scale-dependent portions of higher-order contributions 
to perturbative expressions.  For example, if the two-loop contribution to a physical 
process has been determined via explicit computation of pertinent Feynman diagrams, 
the RGE then determines all leading-log and next-to-leading contributions to {\em all} 
subsequent orders of perturbation theory. We denote such logarithms to be 
``RG-accessible.'' In the present paper we demonstrate how closed-form summation of 
such RG-accessible logarithm contributions is obtained for a number of physical 
processes whose field-theoretical series are known to two or more nonleading orders 
of perturbation theory.

Consider a perturbative series of the form 
\begin{equation}%1
S[x(\mu), L(\mu)] =\sum_{n=0}^\infty x^n S_n\left[xL\right] =
\sum_{n=0}^\infty \sum_{k=0}^n T_{n,k} x^n L^k
\tlabel{pert_series}
\end{equation}
occurring within a physical decay rate $\Gamma$ or measurable 
cross-section $\sigma$, where $x(\mu)$ is the running coupling constant (for
QCD $x(\mu) \equiv \alpha_s (\mu) / \pi$) and where $L(\mu)$ is a
logarithm regulated by the renormalization mass scale $\mu$ that may or
may not also depend on a running mass:
\begin{equation}%2
L(\mu) \equiv \log \left( \frac{\mu^2}{m^2} \right)
\tlabel{log_def}
\end{equation}
If $m$  is a running mass, then
\begin{equation}%3
\mu^2 \frac{\mathrm{d}m}{\mathrm{d}\mu^2}  =  m \gamma_m [x(\mu)]
 =  -m [\gamma_0 x + \gamma_1 x^2 + \gamma_2 x^3 + ...]
\tlabel{mass_RG}
\end{equation}
If $m$ is a pole mass (or for scattering processes, a kinematic
variable), then $\gamma_m$ as defined by \tref{mass_RG} is zero.

For example, in the fully $\overline{{\rm MS}}$ expression for the semileptonic $b
\rightarrow u \ell^- \bar{\nu}_\ell$ rate obtained from 
five active
flavours, 
%(as appropriate for optimization of the rate near $\mu = 2 \,{\rm GeV}$), 
$m$ is the running mass $m_b (\mu)$,
\begin{equation}%4
\Gamma = \frac{G_F^2 |V_{ub}|^2}{192 \pi^3} [m_b (\mu)]^5 S[x(\mu),
L(\mu)]
\tlabel{decay_def}
\end{equation}
and the successive-order series coefficients within $S[x,L]$, 
as defined by \tref{pert_series}, are \cite{vanritbergen}
\begin{equation}%5
T_{0,0} = 1, \; T_{1,0} = 4.25360, \; T_{1,1} = 5, \; T_{2,0} = 26.7848, \; T_{2,1} =
36.9902, \; T_{2,2} = 17.2917
\tlabel{bu_MS_coeff}
\end{equation}
The five active-flavour pole-mass expression for the same rate 
is obtained by replacing $m_b(\mu)$ with the
renormalization-scale independent pole mass $m_b^{pole}$ in \tref{decay_def} and
\tref{log_def}, as well as a concomitant alteration of the following series 
coefficients
\cite{vanritbergen}
\begin{equation}%6
T_{1,0} = -2.41307, \; T_{1,1} = 0, \; T_{2,0} = -21.2955, \; T_{2,1} =
-4.62505, \; T_{2,2} = 0.
\tlabel{bu_pole_coeff}
\end{equation}

Suppose for a given scattering or decay process that the series $S[x,L]$ is known
to some order of perturbation theory:
\begin{gather}%7
S^{NL} = T_{0,0} + \left(T_{1,0} + T_{1,1} L\right)x
\tlabel{SNL_def}
\\
S^{NNL} =
S^{NL}+ \left(T_{2,0} + T_{2,1} L + T_{2,2}
L^2\right) x^2
\tlabel{SNNL_def}
\\
S^{N^3 L}  = S^{NNL}
 + \left(T_{3,0} + T_{3,1} L + T_{3,2} L^2 + T_{3,3} L^3\right)x^3
\tlabel{SN3L_def}
\\
S^{N^4 L} = S^{N^3 L} + \left(T_{4,0} + T_{4,1}L + T_{4,2} L^2 + T_{4,3} L^3
+ T_{4,4} L^4\right) x^4.
\tlabel{SN4L_def}
\end{gather}
These next-to-leading (NL) and higher-order expressions exhibit scale
dependence as the magnitude of $L$ increases.  However, higher order
polynomial coefficients of $L$ can be determined via an appropriate
RGE.  For example, in $b \rightarrow
u \ell^- \bar{\nu}_\ell$ the application of the RGE to the known \cite{
vanritbergen} two-loop (NNL) $\overline{{\rm MS}}$ expression for the rate is
sufficient to determine the three-loop coefficients $T_{3,3}, \;
T_{3,2},$ and $T_{3,1}$:  for $n_f = 5$, $T_{3,3} = 50.914$, $T_{3,2} = 178.76$,
and $T_{3,1} = 249.59$ \cite{ahmady}.  This
procedure is taken a step further in ref.\ \cite{flatness}, in which the four
loop coefficients $T_{4,4}, \; T_{4,3}$, and $T_{4,2}$ are determined via
the RGE for this same process.  Estimates for $T_{3,0}$ are also seen to
determine $T_{4,1}$, yielding an $S^{N^4L}$ expression characterized by
only two unknown coefficients ($T_{3,0}$ and $T_{4,0}$) whose parameter space
can be limited by the constraint \cite{flatness} that successive orders of
perturbation theory decrease in magnitude
\begin{equation}%11
|S^{N^4L} - S^{N^3L}| \lesssim |S^{N^3L} - S^{N^2L}|\lesssim
|S^{N^2L} - S^{NL}|.
\end{equation}

In the present work, we wish to show how {\em all} RG-accessible
logarithms may be summed if $S$ is known to a given order.  Specifically,
we shall obtain explicit all-orders summations for the following four series, as
defined by the intermediate expression in \tref{pert_series}:
\begin{gather}%12
S_0 \left[ x(\mu)L(\mu) \right]  \equiv  T_{0,0} + T_{1,1} xL + T_{2,2}
x^2L^2 + T_{3,3} x^3 L^3 + \ldots
 =   \sum_{n=0}^\infty T_{n,n} x^n L^n
\tlabel{S0_def}
\\
S_1 \left[ x(\mu)L(\mu) \right]  \equiv  T_{1,0} + T_{2,1}
xL + T_{3,2} x^2 L^2 + \ldots
 =  \sum_{n=1}^\infty T_{n,n-1} (x L)^{n-1}
\tlabel{S1_def}
\\
S_2 \left[ x(\mu)L(\mu) \right]  \equiv  T_{2,0} + T_{3,1}
xL + T_{4,2} x^2 L^2 + \ldots
 =  \sum_{n=2}^\infty T_{n,n-2} (x L)^{n-2}
\tlabel{S2_def}
\\
S_3 \left[ x(\mu)L(\mu) \right] \equiv T_{3,0} + T_{4,1}
xL + T_{5,2} x^2 L^2 + \ldots
 =  \sum_{n=3}^\infty T_{n,n-3} (x L)^{n-3}
\tlabel{S3_def}
\end{gather}
The appropriate RGE $\left( \mu^2 \frac{\mathrm{d}}{\mathrm{d}\mu^2} (\Gamma \; \mbox{or}
\; \sigma) = 0\right)$ is seen to determine all series coefficients of
$S_n$ in terms of its leading coefficient $T_{n,0}$, thereby
facilitating the construction of {\em RG-summed} (RG$\Sigma$)
perturbative expressions to any given order of perturbation theory:
\begin{gather}%16
S_{RG\Sigma}^{NL} = S_0 [xL] + x S_1 [xL]
\tlabel{SNL_sum}
\\
S_{RG\Sigma}^{NNL} = S_0 [xL] + x S_1 [xL] + x^2 S_2 [xL]
\tlabel{SNNL_sum}
\\
S_{RG\Sigma}^{N^3L} = S_0 [xL] + x S_1 [xL] + x^2 S_2 [xL] + x^3 S_3 [xL]
\tlabel{SN3L_sum}
\end{gather}
These ${\rm RG}\Sigma$ expressions are seen to exhibit reduced sensitivity to the
renormalization scale $\mu$ even when the logarithms $L$ are quite large.
Compared with the truncated perturbative series,
these resummed expressions more effectively implement the underlying idea
behind the RGE, namely that the exact (all-orders) expression for any physical
quantity is necessarily independent of the scale-parameter $\mu$.

Although RGE determinations of higher-order terms have been known for some time to be of value in extracting divergent parts of bare parameters
\cite{tHooft}, the principle of incorporating {\em all} higher-order RG-accessible 
terms available to a 
given Feynman-diagram order of perturbation theory was, to the best of our knowledge, 
first articulated by Maxwell \cite{maxwell} as a method for eliminating unphysical 
renormalization-scale dependence.  The all-orders summation  of leading logarithms has been
 subsequently applied by Maxwell and 
Mirjalili \cite{mirjalili} to moments of QCD  leptoproduction structure functions 
and to ${\rm NNL}$-order correlation functions. Such a summation of leading-logarithm contributions 
to all orders has also been explicitly performed by McKeon 
to extract one-loop RG functions 
from the effective actions of $\phi^4$-field theory 
in four dimensions and $\phi^6$-field theories in three dimensions \cite{mckeon}. In 
Section \ref{RG_sum_sec} of the present work, we extend McKeon's summation procedure to derive 
closed-form expressions for all-orders  summations of leading- \tref{S0_def}, NL-\tref{S1_def}, 
NNL- \tref{S2_def}, and ${\rm N^3L}$-logarithms \tref{S3_def} by 
using the RGE appropriate to the perturbative series \tref{pert_series} within the QCD expression 
for the inclusive semileptonic B-decay rate.  Such summations enable one to construct 
${\rm RG}\Sigma$ perturbative expressions inclusive of up-to-three nonleading logarithmic 
contributions to all orders of the perturbative series \tref{pert_series}.

In Section \ref{bu_MS_sec}, these results are applied to the $b \rightarrow u\ell^- \bar{\nu}_\ell$ 
rate computed to NNL order by van Ritbergen \cite{vanritbergen}, later extended via 
Pad\'e-approximant methods to a subsequent $N^3 L$ prediction \cite{ahmady}.  
The renormalization-scale dependence of the unsummed perturbative rate truncated 
to a given order is shown to be much greater than that of the ${\rm RG}\Sigma$ rates 
obtained from the same perturbative expression.

In Section \ref{bu_pole_sec}, ${\rm RG}\Sigma$ expressions are obtained for the decays
$b \rightarrow u \ell^- \bar{\nu}_\ell$ and $b \rightarrow c\ell^- \bar{\nu}_\ell$
in the ``pole mass'' scheme in which only the couplant $\alpha_s (\mu)/\pi$ exhibits 
renormalization-scale dependence.  This scheme, already known to have difficulties 
for the $b \rightarrow u$ case \cite{vanritbergen}, exhibits a rate which increases with 
the renormalization scale $\mu$, making the identification of a ``correct'' or optimal 
value of $\mu$ problematical.  However, RG-summation is shown effectively to 
remove such $\mu$-dependence, leading to reliable order-by-order pole-mass-scheme
predictions for the $b \rightarrow u$ semileptonic rate  
consistent with the $b \rightarrow u$ rate obtained from an $\overline{{\rm MS}}$-scheme 
inclusive of a running b-quark mass. ${\rm RG}\Sigma$-expressions are also
obtained for the $b \rightarrow c$ semileptonic rate based upon its
(approximately-) known ${\rm NNL}$ series \cite{czarnecki} and its Pad\'e-estimated 
$N^3 L$ series in the pole-mass scheme \cite{gangbc}.

The RGE appropriate for the perturbative series for semileptonic B-decays 
in the pole mass scheme is also the appropriate RGE for the fermionic vector-current 
correlation function utilised to obtain QCD corrections to the cross-section ratio
$\sigma (e^+ e^- \rightarrow \mbox{hadrons} )/ \sigma (e^+ e^- \rightarrow \mu^+ \mu^-)$.
In Section \ref{vec_corr_sec} we obtain RG-summation expressions for the QCD series embedded within the 
vector-current correlation function that include all higher-order logarithmic contributions 
that are accessible from the three fully-known nonleading orders of perturbative corrections 
in the $\overline{{\rm MS}}$ scheme.  We are thus able to compare directly the renormalization-scale 
dependence of the unsummed series $S^{NL}$ \tref{SNL_def}, $S^{NNL}$ \tref{SNNL_def}, and $S^{N^3 L}$ 
\tref{SN3L_def} 
to their corresponding ${\rm RG}\Sigma$ expressions $S_{RG\Sigma}^{NL}$ \tref{SNL_sum}, 
$S_{RG\Sigma}^{NNL}$ \tref{SNNL_sum}, and $S_{RG\Sigma}^{N^3 L}$ \tref{SN3L_sum}.  We find that the latter 
expressions provide a set of virtually scale-independent order-by-order perturbative 
predictions for the vector correlator.

In Section \ref{pert_sec}, we show how the use of process-appropriate RGE's can be used 
to obtain ${\rm RG}\Sigma$ perturbative expressions for a number of other processes.  We obtain 
full RG-summations for

\begin{enumerate}

\item the momentum-space series for perturbative contributions to the QCD static-potential function,

\item the gluonic scalar-current correlation function characterising scalar gluonium states in QCD sum rules,

\item the (standard-model-) Higgs-mediated cross-section $W_L^+  W_L^- \rightarrow 
Z_L Z_L$ at high energies, which is characterised by the two physical scale parameters
$s$ and $M_H$,

\item the decay of a standard-model Higgs boson into two gluons [a process also characterised by two physical scales 
          ($M_H$ and $M_t$) in addition to the renormalization scale $\mu$], and

\item the fermionic scalar-current correlation function  that characterises both 
          Higgs $\rightarrow b\bar{b}$ decays and scalar-meson-channel QCD sum rules.
\end{enumerate}

\noindent We also discuss how RG-summation of the two scalar-current correlators 
considered in Section \ref{pert_sec} removes much of the unphysical $\mu$-dependence of the unsummed
series at low s that would otherwise percolate through QCD 
sum-rule integrals sensitive to the low-s region.  

In Section 7 we summarise our paper. We discuss not only the reduction of $\mu$-dependence via RG-summation, 
but also the comparison of ${\rm RG}\Sigma$ results with those of unsummed series when minimal 
sensitivity or fastest apparent convergence is used to extract an optimal 
value for the renormalization scale.  

Finally, an alternative all-orders summation procedure to that of Section \ref{RG_sum_sec} is presented in an 
appendix.

\section{RG-Summation of Logarithms for Semileptonic B-Decays}\label{RG_sum_sec}
\renewcommand{\theequation}
{2.\arabic{equation}}
\setcounter{equation}{0}
For semileptonic b-decays, the $\mu$-sensitive portion of the rate \tref{decay_def}
must, as a physically measurable quantity, exhibit renormalization
scale-invariance:
\begin{equation}%2.1
\mu^2 \frac{\mathrm{d}}{\mathrm{d}\mu^2} \left\{ [m_b(\mu)]^5 S[x(\mu),L(\mu)] \right\} =
0.
\end{equation}
This constraint is easily seen to lead to the RGE
\begin{equation}%2.2
\left[1-2\gamma_m (x)\right] \frac{\partial S}{\partial L} + \beta(x)
\frac{\partial S}{\partial x} + 5\gamma_m S = 0,
\tlabel{decay_RGE}
\end{equation}
where
\begin{equation}%2.3
\beta(x) \equiv \mu^2 \frac{\mathrm{d}}{\mathrm{d}\mu^2} x(\mu) = -(\beta_0 x^2 + \beta_1
x^3 + \beta_2 x^4 + \ldots)\quad ,
\tlabel{beta_def}
\end{equation}
$[x(\mu) \equiv \alpha_s (\mu) / \pi]$ and where the anomalous
mass dimension is the series defined by \tref{mass_RG}.  Substitution of the
series expansion \tref{pert_series} into the RGE yields the following series
equation:
\begin{equation}
\begin{split}%2.4
0=&\left(1 +2\gamma_0 x + 2\gamma_1 x^2 + 2\gamma_2 x^3 + \ldots\right) \sum_{n=0}^\infty
\sum_{k=1}^n T_{n,k} k x^n L^{k-1}\\
&- \left(\beta_0 x^2 + \beta_1 x^3 + \beta_2 x^4 + \ldots\right) \sum_{n=1}^\infty
\sum_{k=0}^n T_{n,k} n x^{n-1} L^k
\\
&- 5 \left(\gamma_0 x + \gamma_1 x^2 + \gamma_2 x^3 + \ldots\right) \sum_{n=0}^\infty
\sum_{k=0}^n T_{n,k} x^n L^k \quad .
\end{split}
\tlabel{decay_RG_series}
\end{equation}

\subsection{Evaluation of $S_0$}\label{S0_eval_sec}
To evaluate $S_0[xL]$, as defined by \tref{S0_def}, we use \tref{decay_RG_series} to extract the
aggregate coefficient of $x^n L^{n-1}$ and to obtain the recursion formula
$(n \geq 1)$
\begin{equation}%2.5
n T_{n,n} - \left[\beta_0 (n-1) + 5 \gamma_0\right] T_{n-1, \; n-1} = 0.
\tlabel{S0_decay_recursion}
\end{equation}
We multiply \tref{S0_decay_recursion} by $u^{n-1}$ and sum from $n=1$ to
$\infty$ to obtain the differential equation,
\begin{equation}%2.6
(1-\beta_0 u) \frac{\mathrm{d} S_0 [u]}{\mathrm{d}u} - 5\gamma_0 S_0 [u] = 0,
\tlabel{S0_decay_de}
\end{equation}
where $S_0[u]$ is given by \tref{SNL_sum} with $xL$ replaced by $u$. The solution of 
\tref{S0_decay_de} for the initial condition $S_0 [0] = T_{0,0}$ is
\begin{equation}%2.7
S_0 [u] = T_{0,0}(1 - \beta_0 u)^{-5\gamma_0 / \beta_0}.
\tlabel{S0_decay_res}
\end{equation}
For the special case of pole-mass renormalization schemes $[\gamma_m [x]
= 0]$, $S_0 = T_{0,0} =1$, corresponding to the complete absence of $x^n L^n$
terms from the perturbative series \tref{pert_series} when $n\ge 1$.

\subsection{Evaluation of $S_1$}
To evaluate $S_1[u]$, as defined by \tref{S1_def} with $u = x(\mu) L (\mu)$ we
first extract the aggregate coefficient of $x^n L^{n-2}$ from the RGE
\tref{decay_RG_series} for $n \geq 2$:
\begin{equation}%2.8
\begin{split}
0=&(n-1) T_{n, \; n-1} + 2\gamma_0 (n-1) T_{n-1, \; n-1} - \beta_0 (n-1)
T_{n-1, \; n-2}
\\
&- \beta_1 (n-2) T_{n-2, \; n-2} - 5\gamma_0 T_{n-1, \; n-2} - 5\gamma_1
T_{n-2, \; n-2} \quad.
\end{split}
\tlabel{S1_decay_recursion}
\end{equation}
If one multiplies \tref{S1_decay_recursion} by $u^{n-2}$ and then sums from $n=2$ to
infinity, one obtains the differential equation
\begin{equation}%2.9
(1-\beta_0 u) \frac{\mathrm{d}S_1}{\mathrm{d}u} - (\beta_0 + 5\gamma_0) S_1[u]
= 5\gamma_1 S_0 [u] + (\beta_1 u - 2\gamma_0) \frac{\mathrm{d}S_0}{\mathrm{d}u}.
\tlabel{S1_decay_de}
\end{equation}
We find it convenient to re-express this equation in terms of the
variable
\begin{equation}%2.10
w \equiv 1 - \beta_0 u
\tlabel{w_def}
\end{equation}
and the constant
\begin{equation}%2.11
A \equiv \frac{5\gamma_0}{\beta_0}.
\tlabel{A_def}
\end{equation}
We see from \tref{S0_decay_res} that if $T_{0,0} = 1$, then
\begin{equation}%2.12
S_0 = w^{-A}
\tlabel{S0_w_soln}
\end{equation}
and find from \tref{S1_decay_de} the following differential equation for $S_1$:
\begin{equation}%2.13
\frac{\mathrm{d}S_1}{\mathrm{d}w} + \frac{1+A}{w} S_1 = B w^{-A-1} + C w^{-A-2}
\tlabel{S1_decay_w_de}
\end{equation}
where
\begin{gather}%2.14
B \equiv \left(A \beta_1 - 5\gamma_1\right)/\beta_0
\tlabel{B_def}
\\
C \equiv A\left(2\gamma_0 - \beta_1 / \beta_0\right).
\tlabel{C_def}
\end{gather}
For initial condition $S_1 |_{u=0} = S_1 |_{w=1} = T_{1,0}$, the
solution to \tref{S1_decay_w_de} is
\begin{equation}%2.16
S_1 = Bw^{-A} + \left[T_{1,0} - B + C \log (w)\right]w^{-A-1}
\tlabel{S1_decay_res}
\end{equation}
with $w,A,B$ and $C$ respectively given by \tref{w_def}, \tref{A_def}, \tref{B_def} and
\tref{C_def}.

\subsection{Evaluation of $S_2$}
The aggregate coefficient of $x^n L^{n-3}$ in eq.\ \tref{decay_RG_series} is $(n \geq 3)$
\begin{equation}%2.17
\begin{split}
0=&(n-2) T_{n,n-2} + 2\gamma_0 (n-2) T_{n-1, \; n-2} + 2\gamma_1 (n-2)
T_{n-2, \; n-2}
- \beta_0 (n-1) T_{n-1, \; n-3} 
\\
&- \beta_1 (n-2) T_{n-2, \; n-3} -
\beta_2 (n-3) T_{n-3, \; n-3}
-5\gamma_0 T_{n-1, \; n-3} - 5\gamma_1 T_{n-2, \; n-3} - 5\gamma_2 T_{n-
3, \; n-3} \quad.
\end{split}
\tlabel{S2_decay_recursion}
\end{equation}
If one multiplies \tref{S2_decay_recursion} by $u^{n-3}$ and sums from $n=3$ to
infinity, one finds from the definitions
\begin{gather}%2.18
S_0[u]=1+\sum_{n=1}^\infty T_{n,n} u^n,
\tlabel{S0_decay_def}
\\
S_1[u] = \sum_{n=1}^\infty T_{n, n-1} u^{n-1},
\tlabel{S1_decay_def}
\\
S_2 [u] = \sum_{n=2}^\infty T_{n, n-2} u^{n-2}
\tlabel{S2_decay_def}
\end{gather}
[following from (\ref{S0_def}--\ref{S2_def})] that
\begin{equation}%2.21
\frac{\mathrm{d}S_2}{\mathrm{d}u}  -  \frac{\left(2\beta_0+5\gamma_0\right)}{1-\beta_0 u} S_2
 =  \frac{\left(\beta_1 u - 2\gamma_0\right)}{1-\beta_0 u} \frac{\mathrm{d}S_1}{\mathrm{d}u}
 +  \frac{\left(\beta_2 u - 2\gamma_1\right)}{1-\beta_0 u} \frac{\mathrm{d}S_0}{\mathrm{d}u} +
\frac{\left(\beta_1 + 5\gamma_1\right)}{1-\beta_0 u} S_1
 +  \frac{5\gamma_2}{1-\beta_0 u} S_0\quad.
\tlabel{S2_decay_de}
\end{equation}
If we incorporate the change-of-variable \tref{w_def} in conjunction with the
solutions \tref{S0_w_soln} and \tref{S1_decay_res} for $S_0$ and $S_1$, respectively, we find
that
\begin{equation}%2.22
\frac{\mathrm{d}S_2}{\mathrm{d}w} + \frac{(2+A)}{w} S_2  =  Dw^{-A-1} + E w^{-A-2}
 +  F w^{-A-2} \log (w)
 +  G w^{-A-3} + H w^{-A-3} \log (w)\quad ,
\end{equation}
where the constants $\{D,E,F,G,H\}$ are given by
\begin{gather}%2.23
D = \left[\beta_1 AB + \beta_2 A - \left(\beta_1 + 5\gamma_1\right) B - 5\gamma_2\right] /
\beta_0
\tlabel{D_def}
\\
E  =  \left(2\gamma_0 - \frac{\beta_1}{ \beta_0}\right) AB + 
\left[(1+A)\left(T_{1,0} - B\right) - C\right]
\frac{\beta_1}{\beta_0}
 +  \left(2\gamma_1 - \frac{\beta_2}{\beta_0}\right) A+\left(B - T_{1,0}\right)
\left(\beta_1 +
5\gamma_1\right) / \beta_0
\\
F = \left(A\beta_1 - 5\gamma_1\right)C / \beta_0
\\
G = \left[(1+A)\left(T_{1,0} - B\right) - C\right]\left(2\gamma_0 - \frac{\beta_1}{ \beta_0}\right)
\\
H = \left(2\gamma_0 - \frac{\beta_1}{ \beta_0}\right)(1+A) C
\tlabel{H_def}
\end{gather}
with constants $\{A,B,C\}$ given by \tref{A_def}, \tref{B_def} and \tref{C_def}.  The
solution to the differential equation \tref{S2_decay_de} with initial condition
$S_2|_{u=0} = S_2|_{w=1} = T_{2,0}$ is
\begin{equation}%2.28
\begin{split}
S_2  =&  \frac{D}{2} w^{-A} + (E-F) w^{-A-1} + F w^{-A-1} \log(w)
 +  \left(T_{2,0} - \frac{D}{2} - E+F\right)w^{-A-2} 
\\&+ Gw^{-A-2} \log(w)
 +  \frac{H}{2} w^{-A-2} \log^2 (w).
\end{split}
\tlabel{S2_decay_res}
\end{equation}

\subsection{Evaluation of $S_3$}
The aggregate coefficient of $x^n L^{n-4}$ in \tref{decay_RG_series} is
\begin{equation}%2.29
\begin{split}
0=&(n-3) \left[ T_{n, n-3} + 2\gamma_0 T_{n-1, \; n-3} + 2\gamma_1 T_{n-2,
\; n-3} + 2\gamma_2 T_{n-3, \; n-3} \right]
\\
 &-\beta_0 (n-1) T_{n-1, \; n-4}- \beta_1 (n-2) T_{n-2, \; n-4} - \beta_2
(n-3) T_{n-3, \; n-4}
-\beta_3 (n-4) T_{n-4, \; n-4}
\\
& - 5\gamma_0 T_{n-1, \; n-4} -5\gamma_1
T_{n-2, \; n-4}
-5\gamma_2 T_{n-3, \; n-4} - 5\gamma_3 T_{n-4, \; n-4} \quad.
\end{split}
\tlabel{S3_decay_recursion}
\end{equation}
To evaluate the series
\begin{equation}%2.30
S_3[u] = \sum_{n=3}^\infty T_{n, n-3} u^{n-3}.
\tlabel{S3_decay_def}
\end{equation}
we multiply \tref{S3_decay_recursion} by $u^{n-4}$, sum from $n = 4$ to infinity, and, as before, 
make the eq.\ \tref{w_def} change of variable  $w=1-\beta_0 u$.  We then find that
\begin{equation}%2.31
\begin{split}
\frac{\mathrm{d}S_3}{\mathrm{d}w}  +  \frac{3+A}{w} S_3 
 = & K w^{-A-1} + M w^{-A-2}
 +  N w^{-A-2} \log (w) + P w^{-A-3} + Q w^{-A-3} \log (w)
\\
 &+  R w^{-A-3} \log^2 (w) + U w^{-A-4} + V w^{-A-4} \log (w)
 +  Y w^{-A-4} \log^2 (w)\quad,
\end{split}
\tlabel{S3_decay_de}
\end{equation}
by utilizing the explicit solutions \tref{S0_w_soln}, \tref{S1_decay_res} and \tref{S2_decay_res} for $\{S_0, S_1, S_2 \}$,
as defined by \tref{S0_decay_def}, \tref{S1_decay_def} and \tref{S2_decay_def}.  The new constants within \tref{S3_decay_de} are
\begin{gather}%2.32
K  =  \frac{A}{\beta_0}\left(\beta_3 + B\beta_2 + D \beta_1/2\right)
 - \frac{1}{\beta_0} \left[ 5\gamma_3 + \left(5\gamma_2+\beta_2\right) B + \left(5\gamma_1+2\beta_1\right)
\frac{D}{2}\right] 
\tlabel{K_def}
\\
\begin{split}
M  =&  \left[\left(2\gamma_2 - \frac{\beta_3}{\beta_0}\right) 
+ \left(2\gamma_1 - \frac{\beta_2}{\beta_0}\right)B + \left(2\gamma_0 - \frac{\beta_1}{\beta_0}\right) 
\frac{D}{2} \right] A
\\
&+  \left[ \left(5\gamma_2 + \beta_2\right)\left(B-T_{1,0}\right) + \left(5\gamma_1 +
2\beta_1\right)(F-E)\right] \frac{1}{\beta_0}
\\
 &
 +  \left[ \left(T_{1,0} - B\right)(1+A)-C \right] \frac{\beta_2}{ \beta_0}
 +  \left[ E(1+A) - F(2+A)\right] \frac{\beta_1}{ \beta_0}
\end{split}
\\
N = \left\{ \left(A\beta_2 - 5\gamma_2\right) C+\left[ (A-1)\beta_1 -
5\gamma_1\right] F\right\}\frac{1}{ \beta_0}
\\
\begin{split}
P  =&  \left(2\gamma_1 - \frac{\beta_2}{ \beta_0}\right) \left[ (1+A)\left(T_{1,0} - B\right) -
C\right]
 +  \left(2\gamma_0 - \frac{\beta_1}{ \beta_0}\right) \left[ (1+A) E - (2+A)
F\right]
\\
& - \frac{\left(5\gamma_1 + 2\beta_1\right)}{\beta_0}\left(T_{2,0} - \frac{D}{2} - E + F\right) 
 -  \frac{\beta_1}{\beta_0} \left[ G-(2+A) \left(T_{2,0} - \frac{D}{2} - E + F\right) \right] 
\end{split}
\end{gather}
\begin{gather}
Q  =  \left[ \left(2\gamma_1 - \frac{\beta_2}{ \beta_0}\right) C + \left(2\gamma_0 - 
\frac{\beta_1}{\beta_0}\right) F\right] (1+A)
 -  \left[ \left(5\gamma_1 + 2\beta_1\right) G + (H - (2+A) G) \beta_1 \right] 
\frac{1}{\beta_0}
\\
R = \left(\beta_1 A - 5\gamma_1\right) \frac{H}{ 2\beta_0}
\\
U = \left(2\gamma_0 - \frac{\beta_1}{ \beta_0}\right) \left[ (2+A) \left(T_{2,0} - \frac{D}{2} - E + F\right)
-G \right]
\\
V = \left(2\gamma_0 - \frac{\beta_1}{\beta_0}\right) \left[ (2+A) G - H \right]
\\
Y = \left(2\gamma_0 - \frac{\beta_1}{ \beta_0}\right)(2+A) \frac{H}{2} \quad.
\tlabel{Y_def}
\end{gather}
The constants $\{A,B,C,D,E,F,G,H\}$ within \tref{S3_decay_de} are respectively given by
\{\tref{A_def}, \tref{B_def}, \tref{C_def}, (\ref{D_def}--\ref{H_def})\}.  The solution to \tref{S3_decay_de}, subject to
the initial condition $S_3 |_{w=1} = T_{3,0}$, is
\begin{equation}%2.41
\begin{split}
S_3  =&  \frac{K}{3} w^{-A} + \left(\frac{M}{2} - \frac{N}{4}\right) w^{-A-1}
 +  \frac{N}{2} w^{-A-1} \log (w) + (P-Q + 2R) w^{-A-2}
 +  (Q-2R) w^{-A-2} \log (w) 
\\
&+ R w^{-A-2} \log^2 (w)
 +  \left(-\frac{K}{3} - \frac{M}{2} + \frac{N}{4} - P + Q - 2R + T_{3,0} \right) w^{-A-3}
\\
& +  U w^{-A-3} \log (w) + \frac{V}{2} w^{-A-3} \log^2(w)
 +  \frac{Y}{3} w^{-A-3} \log^3 (w)\quad ,
\end{split}
\tlabel{S3_decay_res}
\end{equation}
where $w=1-\beta_0u$ as in  \tref{w_def}.

\section{Semileptonic  $b \rightarrow u \ell^- \bar{\nu}_\ell$ Decay in the $\overline{{\rm MS}}$ Scheme}%3
\label{bu_MS_sec}
\renewcommand{\theequation}
{3.\arabic{equation}}
\setcounter{equation}{0}
In this section, we consider the semileptonic decay $b \rightarrow u \ell^- \bar{\nu}_\ell$
in the $\overline{\rm{MS}}$ scheme.  The decay rate is given by \tref{decay_def} in
terms of the series \tref{pert_series}.  The coefficients in this series are fully
known to two loop order and are given by \tref{bu_MS_coeff}.  The logarithms $L(\mu)$
within the series are characterized by a running $b$-quark mass, as
given by \tref{log_def} and \tref{mass_RG}.

The coefficients \tref{bu_MS_coeff} are listed for five active flavours, appropriate 
to analysis in an energy region containing $m_b (m_b)$.
%decay rate has been argued \cite{vanritbergen, ahmady} to be optimal at 
%values of $\mu$ substantially below $m_b (m_b)$.  
Consequently, the 
running $b$-quark mass $m_b (\mu)$ and the running couplant $x(\mu) =
\alpha_s (\mu) / \pi$ should be characterized by $n_f = 5$ values for
the RG functions $\gamma_m [x]$ and $\beta[x]$:
\begin{gather}%3.1
\gamma_0 = 1, \; \gamma_1 = \frac{253}{72}, \; \gamma_2 = 7.41986, \; \gamma_3
= 11.0343
\tlabel{anomdim4f}
\\
\beta_0 = \frac{23}{12}, \; \beta_1 = \frac{29}{12}, \; \beta_2 =
\frac{9769}{3456}, \; \beta_3 = 18.8522
\tlabel{beta4f}
\end{gather}
Given the computed values of $T_{1,0}$ and $T_{2,0}$ \cite{vanritbergen},
it is straightforward to calculate the ${\rm NL}$ and ${\rm NNL}$ RG-summations for
the series $S$, as defined in \tref{SNL_sum} and \tref{SNNL_sum}.  The constants $\{A,B,C,\ldots, H\}$
that characterise the summations $S_0, S_1$ and $S_2$ are obtained via
\tref{anomdim4f}, \tref{beta4f} and \tref{bu_MS_coeff} from their definitions in Section \ref{RG_sum_sec}:
\begin{equation}%3.3
\begin{split}
A =& \frac{60}{23}, \; \; B = -\frac{18655}{3174}, \; \; C =\frac{1020}{529} , \; \; D =
26.4461,
\\
E =& -58.8224, \; \; F = -\frac{3171350}{279841}, \; \; G = 25.5973, \; \; H=\frac{1439220}{279841}
\quad .
\end{split}
\tlabel{bu_MS_coeffs}
\end{equation}
Eqs.\ \tref{S0_w_soln}, \tref{S1_decay_res} and \tref{S2_decay_res} then lead to the following closed-form
expressions for the summations $S_0$, $S_1$ and $S_2$:
\begin{gather}%3.4
S_0 = \left[ 1 - \frac{23}{12} x(\mu) L(\mu) \right]^{-60/23}
\tlabel{bu_S0}
\\
\begin{split}%3.5
S_1  = & -\frac{18655}{3174} \left[ 1 - \frac{23}{12} x(\mu) L(\mu) \right]^{-60/23}
\\
 + & \left\{10.1310 + \frac{1020}{529}\log \left[1 - \frac{23}{12} x(\mu) L(\mu)
\right] \right\} \left[ 1 - \frac{23}{12} x(\mu) L(\mu) \right]^{-83/23}
\end{split}
\tlabel{bu_S1}
\\
\begin{split}
S_2  = & 13.2231 \left[ 1 - \frac{23}{12} x(\mu)L(\mu)\right]^{-
60/23}
\\
& -  \left\{47.4897 + \frac{3171350}{279841} \log \left[1-\frac{23}{12} x(\mu) L(\mu)
\right] \right\} \left[1-\frac{23}{12} x(\mu)L(\mu)\right]^{-83/23}
\\
& +  \left\{61.0515+25.5973 \log \left[1-\frac{23}{12}
x(\mu)L(\mu)\right] + \frac{719610}{279841} \log^2 \left[1-\frac{23}{12}
x(\mu)L(\mu)\right] \right\}
 \left[1-\frac{23}{12} x(\mu) L(\mu)\right]^{-106/23}
\end{split}
\tlabel{bu_S2}
\end{gather}

We first wish to compare the $\mu$-dependence of the 2-loop order
expression
\begin{equation}%3.7
\frac{\Gamma^{NNL}}{{\cal K}}   =  \left[m_b (\mu)\right]^5 \left[ 1+(4.25360 + 5
L(\mu)) x(\mu)
 +  \left(26.7848 + 36.9902 L (\mu) + 17.2917 L^2(\mu)
\right)x^2 (\mu) \right]
\tlabel{Gam_MS_NNL}
\end{equation}
for the reduced rate $\left( {\cal K} \equiv G_F^2 |V_{ub}|^2 / 192 \pi^3 \right)$ to that of
the corresponding RG-summed expression
\begin{equation}%3.8
\frac{\Gamma_{RG\Sigma}^{NNL} }{ {\cal K}} = [m_b (\mu)]^5 \left[S_0 + S_1 x(\mu) + S_2 x^2 (\mu) \right]
\tlabel{Gam_sum_form}
\end{equation}
with $S_0,~ S_1$ and $S_2$ given by \tref{bu_S0}, \tref{bu_S1} and \tref{bu_S2}.  To make this comparison, we evolve 
the running coupling and mass from initial values $x(4.17 \,{\rm GeV}) = 0.0715492$ and
$m_b (4.17 \,{\rm GeV}) = 4.17 \,{\rm GeV}$ \cite{ahmady}, 
where the former value arises from $n_f=5$ evolution of the running coupling
from an assumed  anchoring value 
$x\left(M_Z\right)=0.118000/\pi$ \cite{pdg}, and where the latter value is
the $n_f = 5$ central value in ref.\ \cite{chetyrkin} for $m_b (m_b)$.
Thus $x(\mu)$, $m_b (\mu)$ and $L(\mu)$ are fully determined via \tref{log_def}
and the RG-equations \tref{mass_RG} and \tref{beta_def}, with $\gamma_m$- and $\beta$-
function coefficients given by \tref{anomdim4f} and \tref{beta4f}.

\begin{figure}[hbt]
\centering
\includegraphics[scale=0.7]{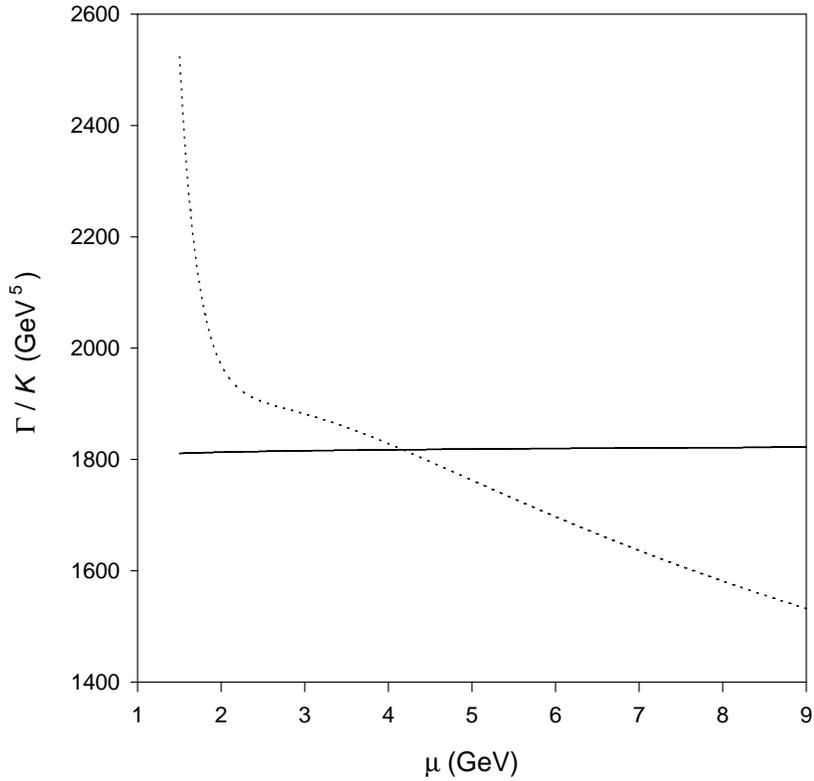}
\caption{
   Comparison of the next-to-next-to-leading- (NNL-) order unsummed (dotted line)
               and RG-summed (solid line) decay rates $\Gamma/{\cal K}$ for 
$b\to u\ell^-\bar\nu_\ell$   in the fully-$\overline{{\rm MS}}$ scheme with
               five active flavours ($n_f = 5$).  The quantity 
${\cal K}\equiv G_F^2\left\vert V_{ub}\right\vert^2/192\pi^3$.
}
\label{resum_f1}
\end{figure}

In Figure \ref{resum_f1}, we use the $n_f = 5$ evolution of $x(\mu)$ and $m_b (\mu)$,
as described above, to compare $\Gamma^{NNL}$ \tref{Gam_MS_NNL} to
$\Gamma_{RG\Sigma}^{NNL}$ \tref{Gam_sum_form}.  It is clear from the figure that
$\Gamma_{RG\Sigma}^{NNL}$ is almost perfectly flat.  By contrast, the
naive rate $\Gamma^{NNL}$ is strikingly dependent on the renormalization
scale $\mu$, and does not exhibit any local extremum point of minimal
sensitivity.  Thus, RG-summation of leading, next-to-leading and 
next-to-next-to-leading logarithms is seen to remove the substantial
theoretical uncertainty associated with the choice of $\mu$ from the
(fully known) two-loop order $b \rightarrow u \ell^- \bar{\nu}_\ell$ rate.

It is useful to examine how the reduced $b \rightarrow u \ell^- \bar{\nu}_\ell$
rate develops in successive orders of perturbation theory.  For example,
the one-loop rates
\begin{gather}%3.9
\frac{\Gamma^{NL}}{{\cal K}} = \left[m_b (\mu)\right]^5 \left[ 1+(4.25360 + 5 L(\mu))
x(\mu)\right]
\tlabel{Gam_MS_NL}
\\
\frac{\Gamma_{RG\Sigma}^{NL}}{ {\cal K}} = \left[m_b (\mu)\right]^5 \left[S_0 + S_1
x(\mu)\right]
\tlabel{Gam_MS_SNL}
\end{gather}
can be compared to the corresponding higher precision results of eq.
\tref{Gam_MS_NNL} and \tref{Gam_sum_form}.  Three-loop order $(N^3 L)$ reduced rates can be
estimated through incorporation of an asymptotic Pad\'e-approximant
prediction of the three-loop coefficient $T_{3,0} = 206$ \cite{ahmady}.
The three-loop order expression for the reduced rate \footnote{Only the 
non-logarithmic three-loop coefficient 206 is estimated; the remaining 
three logarithmic coefficients in (\protect\ref{Gam_MS_N3L}) are obtained via RG-methods in
ref.\ \cite{ahmady}.}
\begin{equation}%3.11
\begin{split}
\frac{\Gamma^{N^3 L}}{ {\cal K}}  =  \left[m_b (\mu)\right]^5 &\Biggl\{ 1 + (4.25360 + 5 L
(\mu)) x(\mu)
+  \left(26.7848 + 36.9902 L(\mu) + 17.2917 L^2 (\mu) \right) x^2(\mu)  \Biggr.
\\
&\quad +  \Biggl. \left( 206 + 249.592 L(\mu) + 178.755 L^2 (\mu) + 50.9144 L^3 (\mu)
\right) x^3 (\mu) \Biggr\}
\end{split}
\tlabel{Gam_MS_N3L}
\end{equation}
can then be compared to its RG-summation version
\begin{equation}%3.12
\frac{\Gamma_{RG\Sigma}^{N^3 L}}{  {\cal K}} = 
\left[m_b (\mu)\right]^5 \left[S_0 + S_1 x(\mu) + S_2 x^2 (\mu) + S_3 x^3 (\mu) \right]
\tlabel{Gam_MS_SN3L}
\end{equation}
with $S_0$, $S_1$ and $S_2$ respectively given by \tref{bu_S0}, \tref{bu_S1} and
\tref{bu_S2}.  The RG-summation $S_3$, is obtained via \tref{S3_decay_res}.  Given the
estimate $T_{3,0} = 206$, the known values \tref{bu_MS_coeffs} of $\{A, B, \ldots, H\}$
and values of $\{K,M,N,P,\ldots,Y\}$ defined via \tref{K_def}--\tref{Y_def},
\begin{equation}%3.13
\begin{split}
K &= -14.3686, \;  M = 146.729, \; N = 50.9925, \;
P = -317.085, \; Q = -148.520, 
\\
  R &=-15.1138 ,\;
U = 189.048, \; V = 83.3941, \; Y = 8.75961,
\end{split}
\end{equation}
we find that
\begin{equation}%3.14
\begin{split}
S_3  = & -\frac{4.7895}{(1-\frac{23}{12} xL)^{60/23}} 
+ \frac{[60.617 + 25.496 \log (1-\frac{23}{12} xL)]}{(1-\frac{23}{12} xL)^{83/23}}
\\
& +  \frac{[-198.79-118.29 \log(1-\frac{23}{12} xL) 
- 15.114 \log^2 (1-\frac{23}{12} xL)]}{(1-\frac{23}{12} xL)^{106/23}} 
\\
& +  \frac{[{\bf 348.96}+189.05 \log (1-\frac{23}{12} xL) + 41.697 \log^2 (1-\frac{23}{12} xL)
+ 2.9199 \log^3 (1 - \frac{23}{12} xL)]}{(1-\frac{23}{12} xL)^{129/23}}
\end{split}
\tlabel{S3_Gam_MS}
\end{equation}
The bold-face number {\bf 348.96} is the only coefficient in the above expression dependent upon
the asymptotic Pad\'e-approximant estimate for $T_{3,0}$.  We have included
this estimate in order to demonstrate RG-summation incorporating a three-loop
diagrammatic contribution to $T_{3,0}$;  when such a calculation is performed, the factor
{\bf 348.96} in \tref{S3_Gam_MS} should then be replaced by $T_{3,0} + 142.96$.

\begin{figure}[hbt]
\centering
\includegraphics[scale=0.7]{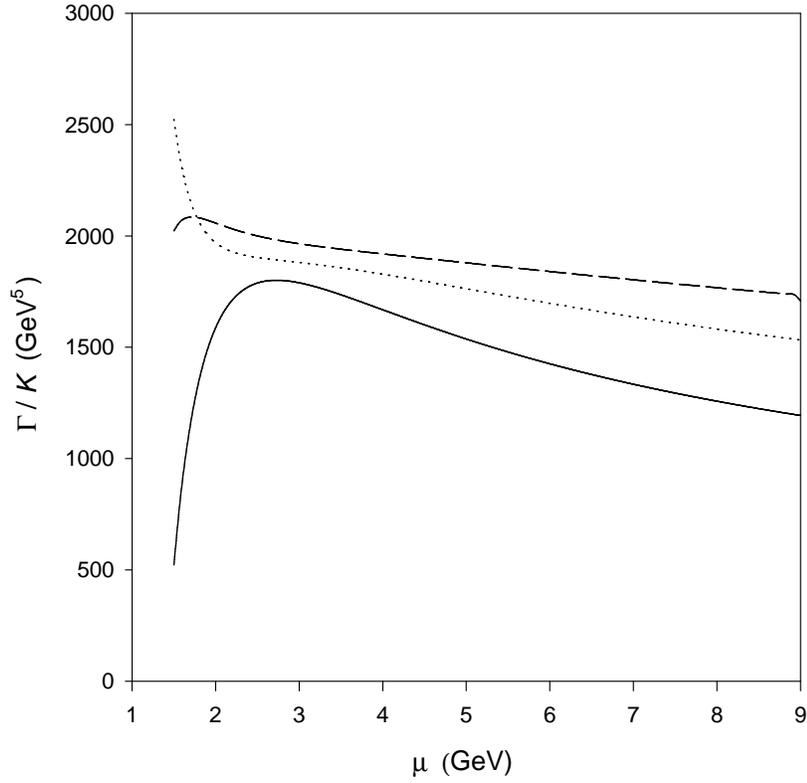}
\caption{
Comparison of unsummed  $b\to u\ell^-\bar\nu_\ell$  decay rates in the fully-$\overline{{\rm MS}}$ scheme 
($n_f =5$) truncated after NL-order (solid line), NNL-order (dotted line), and ${\rm N^3L}$-order
(dashed line).
}
\label{resum_f2}
\end{figure}

We consider the $\mu$-dependence of three non-leading orders of
perturbation theory first for the case in which logarithms are {\em not}
summed to all orders.  Figure \ref{resum_f2} displays a comparison of the
``unsummed'' one-, two- and three-loop order reduced rates $\Gamma^{NL}
/ {\cal K}$, $\Gamma^{NNL} / {\cal K}$ and $\Gamma^{N^3 L} / {\cal K}$, respectively given by
\tref{Gam_MS_NL}, \tref{Gam_MS_NNL} and \tref{Gam_MS_N3L}.  The $\mu$-dependence of all three orders is
evident from the figure. Such $\mu$-dependence can be used to extract ${\rm NL}$
and $N^3 L$ values for $\Gamma$ via the minimal-sensitivity criterion of
ref.\ \cite{stevenson}. Curiously, $\Gamma^{NL}$ and $\Gamma^{N^3 L}$
are both seen to have comparable minimal-sensitivity extrema ($1801\,{\rm GeV}^5$ and
 $2085\,{\rm GeV}^5$) at values of $\mu$ much less than $m_b (m_b)$.
$\Gamma^{NNL}$ exhibits some flattening between these extrema ($\approx
1900 \,{\rm GeV}^5$) over the same range of $\mu$, but with a continued
negative slope. Indeed, one can employ fastest apparent
convergence \cite{grunberg} to choose $\mu$ for $\Gamma^{NNL}$ such that
$|\Gamma^{NNL} (\mu) - \Gamma^{NL} (\mu)|$ is a minimum, and to choose $\mu$
for $\Gamma^{N^3 L}$ such that$|\Gamma^{N^3 L} (\mu) - \Gamma^{NNL} (\mu)| = 0$.  
As evident from Fig.\ \ref{resum_f2}, the former criterion leads to a value for $\mu$ ($2.85 \,{\rm GeV}$) quite 
close to that value at which $\Gamma^{NL}(\mu)$ has an extremum ($2.7 \,{\rm GeV}$), 
corresponding to $\Gamma^{NNL} / {\cal K} = 1888 \,{\rm GeV}^5$.  The latter 
criterion indicates that $\Gamma^{N^3 L}$ should be evaluated at the point 
where $\Gamma^{NNL}$ and $\Gamma^{N^3 L}$ cross, a point noted previously
\cite{ahmady} to be virtually indistinguishable from the 
minimal-sensitivity extremum for $\Gamma^{N^3 L} (\mu)$.

The point we wish to make here, however, is that all such values
extracted for $\mu$ differ substantially from $m_b (m_b)$, in which case
progressively large powers of large-logarithms $L (\mu) \equiv \log
[\mu^2 / m_b^2 (\mu)]$ enter the successive expressions \tref{Gam_MS_NL}, \tref{Gam_MS_NNL} and
\tref{Gam_MS_N3L} for the ${\rm NL}$, ${\rm NNL}$ and $N^3 L$ rate $\Gamma$.  Moreover, in the 
absence of minimal-sensitivity or fastest-apparent-convergence criteria for
extracting $\mu$, even the $N^3 L$  rate exhibits a $\pm 8\%$
spread of values over the range $m_b / 2\lesssim \mu
\stackrel{<}{_\sim} 2m_b$.

\begin{figure}[hbt]
\centering
\includegraphics[scale=0.7]{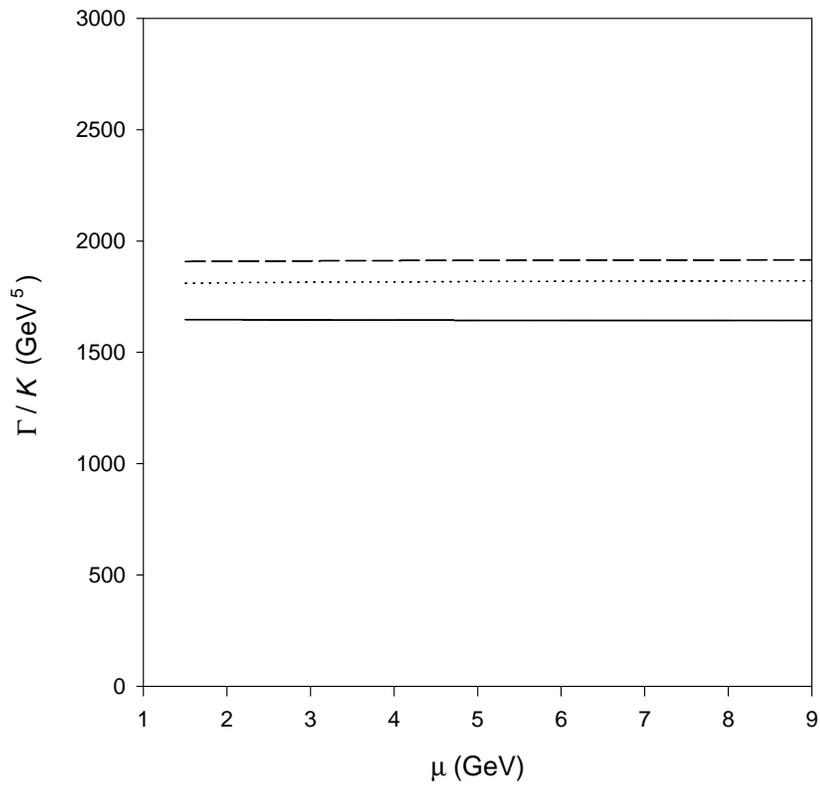}
\caption{Comparison of RG-summation expressions for the fully-$\overline{{\rm MS}}$  $b\to u\ell^-\bar\nu_\ell$
decay rate ($n_f = 5$) obtained from the NL (solid line), NNL (dotted line) and ${\rm N^3L}$ (dashed
line) perturbative series.
}
\label{resum_f3}
\end{figure}

RG-summation eliminates renormalization scale-dependence as a cause of
theoretical uncertainty.  In Figure \ref{resum_f3}, we compare RG-summed versions of
the reduced rate $\Gamma_{RG\Sigma}^{NL}$ \tref{Gam_MS_SNL},
$\Gamma_{RG\Sigma}^{NNL}$ \tref{Gam_sum_form} and $\Gamma_{RG\Sigma}^{N^3 L}$ \tref{Gam_MS_SN3L}.
These three rates exhibit virtually no $\mu$ dependence whatsoever; rather, RG-summation is seen to lead to
clear order-by-order predictions of the rate that are insensitive to $\mu$.
  We see from Fig.\ \ref{resum_f3} that
$\Gamma_{RG\Sigma}^{NL} / {\cal K} = 1646 \pm 2 \,{\rm GeV}^5$,
$\Gamma_{RG\Sigma}^{NNL} / {\cal K} = 1816 \pm 6 \,{\rm GeV}^5$, and
$\Gamma_{RG\Sigma}^{N^3 L} / {\cal K} = 1912 \pm 4 \,{\rm GeV}^5$ over the
(more or less) physical range of $\mu$ considered in Fig.\ \ref{resum_f3}.  Theoretical
uncertainty in the calculated rate is now almost entirely attributable to
truncation of the perturbation series to known contributions, an error
which is seen to diminish as the order of known contributions increases.

It is important to realize, however, that these scale-independent
predictions necessarily coincide with the $L(\mu) = 0$ predictions of
the unsummed rates \tref{Gam_MS_NL}, \tref{Gam_MS_NNL} and \tref{Gam_MS_N3L}:  $\Gamma^{N^k L}$ and
$\Gamma_{RG\Sigma}^{N^k L}$ equilibrate when $w = 1$ {\it i.e.}, when
$L(\mu) = 0$.  Thus, one can argue that the summation we have performed
here of all RG-accessible logarithms supports the prescription of
identifying as ``physical'' those perturbative results in which 
$\mu$-sensitive logarithms are set equal to zero.  We must nevertheless 
recognize the possibility that the $\mu$-sensitivity of the unsummed rates, 
when exploited by minimal-sensitivity or fastest-apparent-convergence 
criteria, is capable of leading to more accurate order-by-order 
estimates of the true rate than corresponding scale-independent
${\rm RG}\Sigma$ rates.  In comparing Figs.\ \ref{resum_f2} and \ref{resum_f3}, it is
noteworthy that the $1801 \,{\rm GeV}^5$ extremum of $\Gamma^{NL} /
{\cal K}$, the {\em unsummed} one-loop rate, is quite close to
$\Gamma_{RG\Sigma}^{NNL} / {\cal K}$, the RG-summed {\em two-loop} rate.  Similarly,
the $1888 \,{\rm GeV}^5$ fastest-apparent-convergence value  of
$\Gamma^{NNL} / {\cal K}$, the unsummed two-loop rate, is close to
$\Gamma_{RG\Sigma}^{N^3 L}$.  This train of argument would suggest that
the ${\cal{O}}(2085 \,{\rm GeV}^5)$ extremum (or fastest-apparent-convergence
value) of the unsummed rate $\Gamma^{N^3 L} / {\cal K}$ may be a more accurate
estimate of the {\em true} rate than $\Gamma_{RG\Sigma}^{N^3 L}$.  Such
an argument, however, requires substantiation via explicit three- and
four-loop order calculations, computations which are not yet available.

\section{Application to Semileptonic B decays in the Pole-Mass Scheme}
\label{bu_pole_sec}
\renewcommand{\theequation}
{4.\arabic{equation}}
\setcounter{equation}{0}
In the pole-mass renormalization scheme, the mass $m$ appearing in
logarithms \tref{log_def} is independent of the renormalization mass-scale $\mu$.
Thus the coefficients $\gamma_k$, as defined in \tref{mass_RG} are all zero.  
The constants $\{A,B,C,D,E,F,H,$ $K,M,N,Q,R,Y\}$, as defined in
Section \ref{RG_sum_sec}, are all zero as well. The nonzero constants are
\begin{gather}%4.1
G = -T_{1,0} \frac{\beta_1}{ \beta_0}
\tlabel{G_pole_res}
\\
P = -T_{1,0} \left(\beta_2 - \frac{\beta_1^2 }{ \beta_0}\right)\frac{1}{ \beta_0}
\\
U =\left[ -T_{1,0} \frac{\beta_1 }{ \beta_0} - 2 T_{2,0} \right] \frac{\beta_1}{\beta_0}
\\
V = 2 T_{1,0} \frac{\beta_1^2}{ \beta_0^2}
\tlabel{V_pole_res}
\end{gather}
and the corresponding RGE summations appearing within $S_{RG\Sigma}^{NL},
S_{RG\Sigma}^{NNL}$, and $S_{RG\Sigma}^{N^3L}$ [\tref{SNL_sum}, \tref{SNNL_sum} and \tref{SN3L_sum}] are
\begin{gather}%4.5
S_0 = 1
\tlabel{S0_pole}
\\
S_1 [xL] = T_{1,0} / (1 - \beta_0 xL)
\tlabel{S1_pole}
\\
S_2 [xL] = T_{2,0} (1 - \beta_0 xL)^{-2} + G(1-\beta_0 xL)^{-2} \log (1-
\beta_0 xL)
\tlabel{S2_pole}
\\
S_3 [xL]  =  P(1-\beta_0 xL)^{-2}
 +  \left[ (T_{3,0}-P)+U\log(1-\beta_0 xL) 
+  \frac{V}{2} \log^2 (1-\beta_0 xL) \right] (1 - \beta_0 xL)^{-3}\quad.
\tlabel{S3_pole}
\end{gather}
In this section, we apply the above results toward the decay $b
\rightarrow u \ell^- \bar{\nu}_\ell$ and $b \rightarrow c \ell^-
\bar{\nu}_\ell$.  The former rate is known fully to two-loop order in the
pole mass scheme, though the result is argued to be of limited phenomenological
utility \cite{vanritbergen}.  The latter rate has been estimated within fairly
narrow errors to two-loop order as well \cite{czarnecki}, and has been extended to a
three-loop order estimated rate via asymptotic Pad\'e-approximate
methods \cite{gangbc}.

\subsection{Pole Scheme Semileptonic $b \rightarrow u \ell^- \bar{\nu}_\ell$ Decay}
\begin{figure}[hbt]
\centering
\includegraphics[scale=0.7]{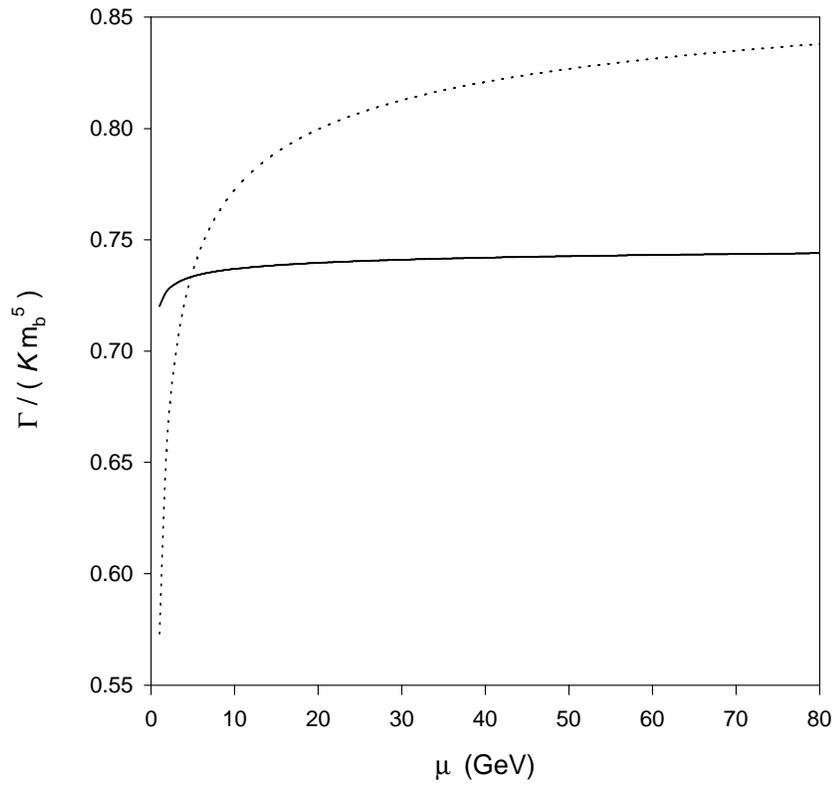}
\caption{
Comparison of the large-$\mu$ behaviour of the NNL unsummed (dotted line) and
               RG-summed (solid line) decay rates for  $b\to u\ell^-\bar\nu_\ell$  within the RG-invariant
               pole-mass scheme with five active flavours.
}
\label{resum_f4}
\end{figure}

The two-loop order $b \rightarrow u \ell^- \bar{\nu}_\ell$ rate in the
pole mass scheme is given by substitution of known values of the
coefficients $\{T_{1,0}, T_{1,1}, T_{2,0}, T_{2,1}, T_{2,2}\}$, as
listed in \tref{bu_pole_coeff} into the series \tref{decay_def}, with $m_b (\mu)$ replaced by
$m_b^{pole}$ as noted earlier.  Although the reliability of the 
pole-mass scheme for this process is suspect because of the proximity of a
renormalon pole \cite{vanritbergen}, we have plotted this series for a range of
$x(\mu) = \alpha_s (\mu) / \pi$ and a choice for $m_b^{pole}$ that will
facilitate comparison with phenomenology already obtained from the
corresponding $\overline{\rm{MS}}$ process.  We choose $n_f = 5$ active
flavours in order to explore the $\mu$-dependence of the NNL rate in a
region in which $\mu$ is considerably larger than $m_b^{pole}$.
Corresponding results for four active flavours are easily obtainable as
well.  The evolution of $x(\mu) = \alpha_s (\mu) / \pi$ for five active
flavours ultimately devolves from $\alpha_s (M_z) = 0.118$ and leads to
the same $n_f = 5$ benchmark value $x(4.17 \,{\rm GeV}) = 0.071549$ as 
noted in Section \ref{bu_MS_sec}.  Similarly we employ a value
for $m_b^{pole} = 4.7659$ consistent to two-loop order with our use of
the running mass value $m_b (m_b) = 4.17 \,{\rm GeV}$, as obtained from the
$n_f = 5$ relation between $m_b^{pole}$ and $m_b (m_b)$ of refs.\ \cite{chetyrkin, pole}:
\begin{equation}%4.9
m_b^{pole} = 4.17 \,{\rm GeV} \left[ 1 + \frac{4}{3}x (4.17) + 9.27793 x^2
(4.17) \right]
\tlabel{mb_pole}
\end{equation}
We then see from Fig.\ \ref{resum_f4} that the $\mu$-sensitive portion of the known
two-loop rate in the pole mass scheme,
\begin{equation}%4.10
S^{NNL} (\mu) = 1 - 2.41307 x(\mu) + \left[-21.2955-4.62505 
\log\left\{ \left(\frac{\mu}{m_b^{pole}}\right)^2\right\}\;\right] x^2 (\mu)
\end{equation}
is indeed highly scale dependent.  Specifically, we see that
$S^{NNL}(\mu)$ increases monotonically with $\mu$ without exhibiting an
extremum identifiable with a ``physical'' point of minimal sensitivity
\cite{stevenson}.

In Figure \ref{resum_f4} we have also plotted the RG-summed version of the 2-loop rate
\begin{equation}%4.11
S_{RG\Sigma}^{NNL}  =  1 + x(\mu) S_1\left[x(\mu)
\log\left\{ \left(\frac{\mu}{m_b^{pole}}\right)^2\right\}
\right]
 +  x^2(\mu) S_2\left[x(\mu)\log\left\{ \left(\frac{\mu}{m_b^{pole}}\right)^2\right\}\right]
\end{equation}
with $x(\mu)$ and $m_b^{pole}$ as obtained above. The summations $S_1$
and $S_2$ are obtained via \tref{S1_pole} and \tref{S2_pole} using the $n_f = 5$ pole-mass
scheme values $T_{1,0} = -2.41307$, $T_{2,0} = -21.2955$
\cite{vanritbergen}
and $n_f = 5$ QCD $\beta$-function coefficients $\beta_0 = 23/12$,
$\beta_1 = 29/12$, and $\beta_2 = 9769/3456$.  It is evident from the
figure that renormalization scale-dependence is considerably reduced by
the summation of all orders of leading and next-to-leading logarithms in
\tref{Gam_MS_N3L}.  The increase of $S_{RG\Sigma}^{NNL}$ with increasing $\mu$ is
minimal compared to that of $S^{NNL}(\mu)$, the unsummed expression.

\begin{figure}[hbt]
\centering
\includegraphics[scale=0.7]{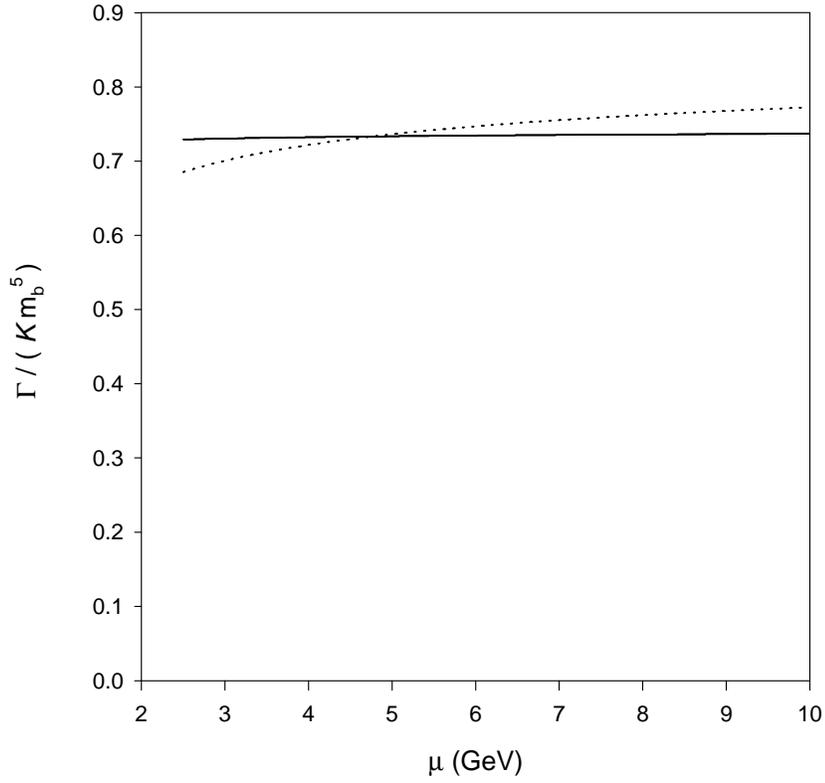}
\caption{
   Comparison of the NNL unsummed (dotted line) and RG-summed (solid line)
               decay rates for    $b\to u\ell^-\bar\nu_\ell$  ($n_f = 5$) within the pole-mass scheme over the
               range $m_b^{pole}/2\lesssim \mu\lesssim 2m_b^{pole}$.
}
\label{resum_f5}
\end{figure}

In Figure \ref{resum_f5}, the comparison between $(m_b^{pole})^5  S^{NNL}$ and 
$ (m_b^{pole})^5 S_{RG\Sigma}^{NNL}$ is
exhibited over the physically relevant region $m_b^{pole}/2
\stackrel{<}{_\sim} \mu\lesssim 2 m_b^{pole}$.  The crossing
point between the two curves necessarily occurs when $L(\mu) = 0$,
corresponding to $\mu = m_b^{pole}$.  Since $S_{RG\Sigma}^{NNL}$ is
insensitive to $\mu$, this crossover supports the expectation discussed 
in the previous section that the ``physical'' ${\rm NNL}$ rate is $S^{NNL}$ with $\mu$
chosen to make all logarithms vanish.  We would prefer, however, to
argue that $S_{RG\Sigma}^{NNL}$ is an almost scale-independent formulation of
the ${\rm NNL}$ rate, thereby obviating any need to define a physically
appropriate value of $\mu$ to compute a meaningful two-loop order
result.  We also note that the asymptotic large-$\mu$ result we obtain
for the ``reduced rate''
\begin{equation}%4.12
\frac{\Gamma_{RG\Sigma}^{NNL}}{  {\cal K}} \cong (m_b^{pole})^5 S_{RG\Sigma}^{NNL} 
\begin{array}{c}{}\\
\longrightarrow\\ ^{\mu = M_W} \end{array}
  1829 \,{\rm GeV}^5
\end{equation}
is surprising close to the $1817 \,{\rm GeV}^5$ $\overline{\rm{MS}}$ two-loop
order (``unsummed'' NNL) estimate obtained at $\mu = m_b (\mu) = 4.17 \,{\rm GeV}$, indicative
of the utility of the pole-mass scheme when leading and next-to-leading
logarithms are summed to all orders. In the absence of such summation
the pole mass expression $S^{NNL} (\mu)$ spans values for the reduced
rate between $1420 \,{\rm GeV}^5$ and $2060 \,{\rm GeV}^5$ as $\mu$
increases from $1 \,{\rm GeV}$ to $M_W$, reflecting the problems with the
pole-mass scheme already noted in ref.\ \cite{vanritbergen}. 
By contrast, the RG-summed reduced rate varies
only from $1774 \,{\rm GeV}^5$ to $1829 \,{\rm GeV}^5$ over the same region of
$\mu$.

\subsection{Pole Scheme Semileptonic $ b \rightarrow c \ell^- \bar{\nu}_\ell$ Decay}
The semileptonic decay of $B$ into a charmed hadronic state is given by the
following decay rate in the pole-mass renormalization scheme
\cite{czarnecki}:
\begin{equation}%4.13
\Gamma(b \rightarrow c \ell^- \bar{\nu}_\ell) =
\frac{G_F^2|V_{cb}|^2}{192\pi^3} F\left(\frac{m_c^2}{m_b^2}\right) m_b^5
S\left[x(\mu), L(\mu)\right]
\tlabel{Gam_bc}
\end{equation}
In \tref{Gam_bc}, $m_b$ and $m_c$ are $\mu$-invariant pole masses, $L(\mu)
\equiv \log\left(\mu^2 / m_b m_c\right)$, $x(\mu)=\alpha_s(\mu)/\pi$, and
$F(r)$ is the form factor
\begin{equation}%4.14
F(r) = 1 - 8r - 12r^2 \log(r) + 8r^3 - r^4 \; .
\end{equation}
{\em All} sensitivity to the renormalization scale $\mu$ resides in the series
$S[x,L]$, which may be expressed in the usual form
\begin{equation}%4.15
S[x,L]  =  1+\left(T_{1,0} + T_{1,1} L\right) x + \left(T_{2,0} + T_{2,1} L + T_{2,2}
L^2\right)
x^2
 +  \left(T_{3,0} + T_{3,1} L + T_{3,2} L^2 + T_{3,3} L^3\right) x^3
+\ldots \quad .
\tlabel{S_bc}
\end{equation}
For four active flavours, the perturbatively calculated coefficients of
\tref{S_bc} are $T_{1,0} = -1.67$ and the (partially-estimated) coefficient
$T_{2,0} = -8.9(\pm 0.3)$ \cite{czarnecki, armadillo}.  Except for
$T_{3,0}$, the remaining coefficients in \tref{S_bc} are accessible from the RG
equation
\begin{equation}%4.16
0 = \left[ \frac{\partial}{\partial L} + \beta (x)
\frac{\partial}{\partial x} \right] S[x,L] \; .
\tlabel{bc_RGE}
\end{equation}
These coefficients are $T_{n,n} = 0 \; [n\geq 1], \; T_{2,1} = T_{1,0} \beta_0
= -3.479$, $T_{3,1} = 2T_{2,0} \beta_0 + T_{1,0} \beta_1 = -42.4(\pm 1.3)$, and
$T_{3,2} = T_{1,0} \beta_0^2 = -7.25$ \cite{gangbc}.  An asymptotic 
Pad\'e-approximant estimate of $T_{3,0} = -50.1 (\pm 2.6)$ has also been
obtained in ref.\ \cite{gangbc}.  Consequently, one may list three orders
for the $\mu$-dependent portion of the $b \rightarrow c \ell^- \bar{\nu}_\ell$ rate:
\begin{gather}%4.17
S^{NL}\left[x(\mu),L(\mu)\right] = 1 - 1.67 x(\mu),
\tlabel{S_bc_NL}
\\
S^{NNL}[x(\mu),L(\mu)]=1-1.67 x(\mu) + \left[-8.9 - 3.479 L(\mu)\right] x^2 (\mu),
\tlabel{S_bc_NNL}
\\
S^{N^3L} [x(\mu), L(\mu)] = S^{NNL} [x(\mu), L(\mu)] + \left[-50.1 -
42.4L(\mu) - 7.25L^2 (\mu)\right] x^3 (\mu),
\tlabel{S_bc_N3L}
\end{gather}
with the caveat that ${\rm NNL}$ and ${\rm N^3L}$ expressions have increasing theoretical uncertainty
arising from the (small) estimated error in $T_{2,0}$ and 
concomitant error in the estimation of $T_{3,0}$.  

\begin{figure}[hbt]
\centering
\includegraphics[scale=0.7]{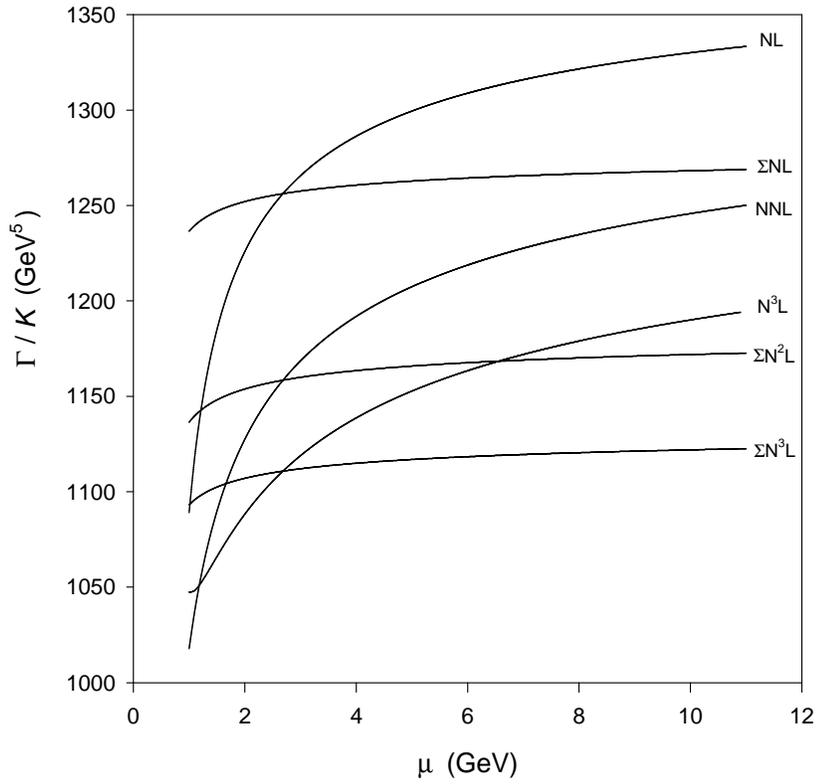}
\caption{
  Comparison of unsummed  and RG-summed  $b\to c\ell^-\bar\nu_\ell$ decay rates $\Gamma/{\cal K}$ ($n_f = 4$) in the  pole-mass scheme, where  ${\cal K}\equiv G_F^2\left\vert V_{cb}\right\vert^2/192\pi^3$.
The curves representing the unsummed rates are labelled 
by NL, NNL and
${\rm N^3L}$ indicating the order at which they are truncated.  Similiarly, the RG-summed 
curves are labelled by $\Sigma{\rm NL}$, $\Sigma{\rm N^2L}$ and $\Sigma{\rm N^3L}$. 
}
\label{resum_f6}
\end{figure}

As before, we will
compare the $\mu$ dependence of \tref{S_bc_NL}, \tref{S_bc_NNL} and \tref{S_bc_N3L} to 
that of the corresponding ${\rm RG}$-summed expressions
\begin{gather}%4.20
S_{RG\Sigma}^{NL} = 1 + x(\mu) S_1\left[x(\mu) L(\mu)\right]
\tlabel{S_bc_SNL}
\\
S_{RG\Sigma}^{NNL}  =  1+ x(\mu) S_1\left[x(\mu)L(\mu)\right]
 +  x^2 (\mu) S_2\left[x(\mu)L(\mu)\right]
\tlabel{S_bc_SNNL}
\\
S_{RG\Sigma}^{N^3L} = S_{RG\Sigma}^{NNL} + x^3 (\mu) S_3\left[x(\mu)L(\mu)\right],
\tlabel{S_bc_SN3L}
\end{gather}
in order to illustrate how ${\rm RG}$-summation of higher logarithms affects
the order-by-order renormalization-scale dependence of a perturbative
series.
Figure \ref{resum_f6} displays a plot of the $\mu$-sensitive portions 
of the decay rate considered to ${\rm NL}$ \tref{S_bc_NL}, ${\rm NNL}$ \tref{S_bc_NNL} and ${\rm N^3L}$
\tref{S_bc_N3L} orders.  As in \cite{czarnecki, gangbc}, $m_c$ is assumed to be
$m_b / 3$.  We have chosen the pole mass $m_b$ to be $4.9 \,{\rm GeV}$
consistent with phenomenological estimates \cite{hoang}.\footnote{Such 
an estimate is slightly larger than that based upon
\tref{mb_pole}, as \tref{mb_pole} is truncated after two-loop order.}  The couplant
$x(\mu) = \alpha_s (\mu)/\pi$ is chosen to devolve from $\alpha_s
(m_\tau) = 0.33$ \cite{alpha} via four active flavours,  
where $\beta_0=25/12$, $\beta_1=77/24$, $\beta_2=21943/3456$ and $\beta_3=31.38745$.
These choices permit careful
attention to the $1.5 \,{\rm GeV}\lesssim \mu\lesssim
m_b$ low-scale region anticipated to correspond to the physical rate,
although we have chosen to extend the range of $\mu$ to $\sim 2m_b$ in
Fig.\ \ref{resum_f6}.  

Figure \ref{resum_f6} demonstrates that the rate expressions appear
to progressively flatten with the inclusion of higher order corrections,
but that the residual scale dependence of each order remains comparable
to the difference between successive orders. Figure \ref{resum_f6} also displays rates 
proportional to the corresponding RG-summed expressions
\tref{S_bc_SNL}, \tref{S_bc_SNNL} and \tref{S_bc_SN3L} based upon the same phenomenological inputs.  
The expressions
for $S_1$, $S_2$ and $S_3$ are given by \tref{S1_pole}, \tref{S2_pole} and \tref{S3_pole}.  It is
evident from Figure \ref{resum_f6} that the scale dependence of RG-summed expressions
to a given order is dramatically reduced from the scale dependence of
the corresponding non-summed expressions.

Thus $S_{RG\Sigma}^{NL}$, $S_{RG\Sigma}^{NNL}$ and $S_{RG\Sigma}^{N^3L}$
are effective scale-independent formulations of the one-, two- and
three-loop perturbative series within the $b \rightarrow c \ell^-
\bar{\nu}_\ell$ rate;  once again the summation of progressively less-than-leading
logarithms in the perturbative series is seen to remove the choice of
renormalization-scale $\mu$ as a source of theoretical uncertainty to
any given order of perturbation theory.

\section{The Vector-Current Correlation Function and $R(s)$}\label{vec_corr_sec}
\renewcommand{\theequation}
{5.\arabic{equation}}
\setcounter{equation}{0}
The imaginary part of the $\overline{\rm{MS}}$ vector-current correlation function for massless quarks can be
extracted from the Adler function \cite{gorishny}.  This procedure is
explicitly given in \cite{ces} and leads to an expression in the
following form:
\begin{equation}%5.1
\frac{1}{\pi} {\rm Im}\; \Pi_v (s) = -\frac{4}{3} \sum_f Q_f^2 S\left[x(\mu), \log (\mu^2 / s)\right],
\tlabel{Im_Pi_v}
\end{equation}
where $x(\mu) = \alpha_s (\mu)/\pi$ and $s$ is the kinematic variable $p^2$
[{\it i.e.} the square of the invariant mass in $e^+ e^- \rightarrow$ hadrons].  The series
$S[x, L]$ appearing in \tref{Im_Pi_v} is fully-known to $N^3 L$ order:
\begin{equation}%5.2
S^{N^3 L} [x, L]  =  1 + x + (T_{2,0} + T_{2,1} L) x^2
 + \left(T_{3,0} + T_{3,1} L + T_{3,2} L^2\right) x^3 .
\tlabel{S_vec_N3L}
\end{equation}
The full series is, of course, identifiable with the generic series \tref{pert_series} [or \tref{S_bc}] 
provided $T_{0,0} = T_{1,0} = 1$ and $T_{n,n} = 0 \; [n \geq 1]$.  
Values for the remaining vector correlation function constants in \tref{S_vec_N3L}
are tabulated in Table \ref{resum_tab1} for three, four and five flavours.  The
coefficients $T_{2,0}$ and $T_{3,0}$ are obtained from the results of
ref.\ \cite{gorishny}, and are well known from the standard expression
for perturbative contributions to $R(s)$.  The remaining coefficients
\begin{equation}%5.3
T_{2,1} = \beta_0, \; \; T_{3,1} = 2\beta_0 T_{2,0} + \beta_1, \; \;
T_{3,2} = \beta_0^2
\end{equation}
are easily determined from the renormalization scale-invariance of the
vector-current correlation function \tref{Im_Pi_v},
\begin{equation}%5.4
\mu^2 \frac{\mathrm{d}}{\mathrm{d}\mu^2} S[x(\mu), \; \log (\mu^2 / s)] = \left[
\frac{\partial}{\partial L} + \beta(x) \frac{\partial}{\partial x}
\right] S[x, L] = 0
\tlabel{RG_vec}
\end{equation}
This equation, of course, can be interpreted to reflect the
imperviousness of the physical quantity
\begin{equation}%5.5
R(s) \equiv \frac{\sigma(e^+ e^- \rightarrow \mbox{hadrons})}{\sigma(e^+
e^- \rightarrow \mu^+ \mu^-)} = -\frac{3}{4\pi} {\rm Im}\; \Pi_v (s),
\end{equation}
to changes in the choice of QCD renormalization-scale $\mu$ \cite{yndurain}.

Equation \tref{RG_vec} is just the RGE \tref{decay_RGE} with $\gamma_m (x)$ set equal to
zero--- precisely the same RG-equation as applicable to the pole-mass
scheme semileptonic $b$ decay rates considered in the previous section.
Consequently, the RG-summation of the series $S[x, L]$ within \tref{Im_Pi_v}
involves the {\em same} series summations $S_0, \; S_1, \; S_2$ and $S_3$
as those given by \tref{S0_pole}, \tref{S1_pole}, \tref{S2_pole} and \tref{S3_pole}.  The (nonzero) constants $G,
\; P, \; U$ and $V$ appearing in these equations are found in terms of
$\beta$-function coefficients $\beta_0, \; \beta_1, \; \beta_2$ and
series coefficients $T_{1,0} (=1)$, $T_{2,0}$ and $T_{3,0}$ via eqs.\ \tref{G_pole_res} --\tref{V_pole_res}.
These constants are all tabulated in Table \ref{resum_tab1}, and are seen to fully
determine the ${\cal{O}}(N^3 L)$ RG-summed version of the series $S[x,
L]$,
\begin{equation}%5.6
S_{RG\Sigma}^{N^3 L} [ x, L] = 1 + x
S_1 [x L] + x^2 S_2 [x L] + x^3 S_3 [x L] .
\tlabel{S_vec_SN3L}
\end{equation}
where $x = x(\mu)$ and $L = \log\left(\mu^2 / s\right)$.  For example, if $n_f = 5$, we see 
from \tref{S1_pole}, \tref{S2_pole} and \tref{S3_pole} and the
Table \ref{resum_tab1} entries for $G, P, U, V, T_{2,0}$ and $T_{3,0}$ that
\begin{gather}%5.7
S_1 [x L] = \frac{1}{ \left( 1 - \frac{23}{12} x L \right)}
\\
S_2 [x L] = \frac{1.49024 - 1.26087 \log \left( 1 - \frac{23}{12} x L
\right)}{\left( 1 - \frac{23}{12} x L\right)^2}
\\
S_3 [x L]  =  \frac{0.115003}{\left( 1 - \frac{23}{12} x L \right)^2}
 +  \frac{\left[-12.9196 - 5.14353 \log \left( 1 - \frac{23}{12} x L
\right) + 1.58979 \log^2 \left( 1 - \frac{23}{12} x L \right) \right]}
{\left( 1 - \frac{23}{12} x L \right)^3}
\end{gather}

\begin{figure}[hbt]
\centering
\includegraphics[scale=0.7]{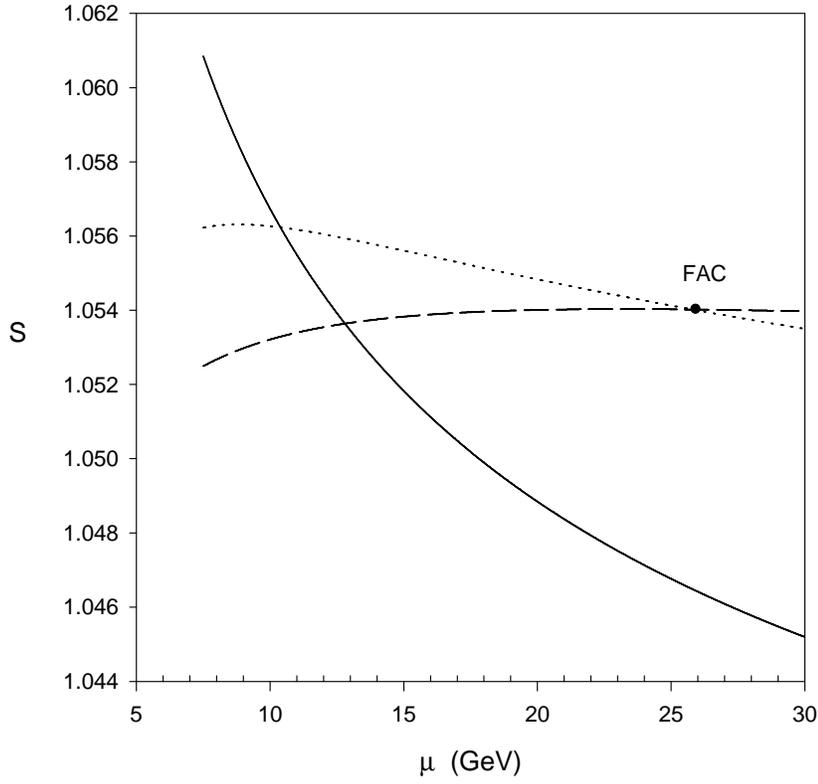}
\caption{
    Comparison of the following $n_f = 5$ vector-current correlation-function series
               when  $\sqrt{ s} = 15  \,{\rm GeV}$: $S^{NL}$ (solid line), the $\overline{{\rm MS}}$ 
perturbative series $S\left[x(\mu   ), \log\left(\mu^2 /s\right)\right]$
               truncated after NL-order contributions; $S^{NNL}$ (dotted line), the same series
               truncated after NNL-order contributions; and $S^{N^3L}$      (dashed line), the same series
               truncated after ${\rm N^3L}$-order contributions. At the intersection of 
                $S^{NNL}$ with $S^{N^3L}$, the
               ${\rm N^3L}$-order contribution to the perturbative series is zero, corresponding to the
               point of fastest apparent convergence (FAC).
}
\label{resum_f8}
\end{figure}

\begin{figure}[hbt]
\centering
\includegraphics[scale=0.7]{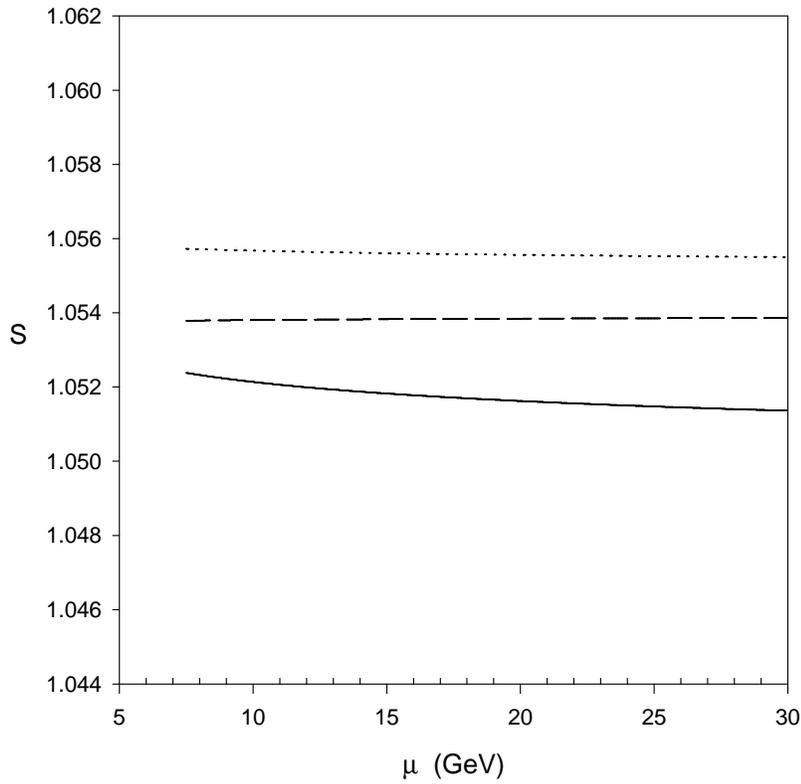}
\caption{
      Comparison of the following RG-summations of the perturbative $\overline{{\rm MS}}$ series
               within the $n_f = 5$ vector-current correlation-function when   $\sqrt{ s} = 15 \,{\rm GeV}$: 
$S_{RG\Sigma}^{NL}$ 
               (solid line), the summation based upon the perturbative series 
$S\left[x(\mu   ), \log\left(\mu^2 /s\right)\right]$
               truncated after NL-order contributions; 
 $S_{RG\Sigma}^{NNL}$ (dotted line), the summation based
               upon the perturbative series truncated after NNL-order contributions; and 
 $S_{RG\Sigma}^{N^3L}$      
               (dashed line), the summation based upon the perturbative series truncated after
                ${\rm N^3L}$-order contributions.
}
\label{resum_f9}
\end{figure}

In Figures \ref{resum_f8} and \ref{resum_f9}, we compare the $\mu$-dependence of the unsummed \tref{S_vec_N3L} and summed \tref{S_vec_SN3L}
expressions for $S[x(\mu), \log (\mu^2 / s)]$, with the choice $s=(15
\,{\rm GeV})^2$.  The running coupling constant $x(\mu)$ is assumed to evolve
via the $n_f = 5$ (four-loop-order) $\beta$-function from an initial
value $x(M_z) = 0.11800/\pi$.\footnote{For purposes of comparing the
$\mu$-dependence of $S^{N^3 L}$ and $S_{RG\Sigma}^{N^3 L}$, we are
assuming (as in Sections \protect\ref{bu_MS_sec} and \protect\ref{bu_pole_sec}) there to be no uncertainty in the value of $\alpha_s (\mu)$.}
Although Figure \ref{resum_f8} does show a flattening of the unsummed expressions upon incorporation
of successively higher orders of perturbation theory $[S^{NL}, S^{NNL},
S^{N^3 L}]$, Figure \ref{resum_f9} demonstrates that the corresponding RG-summed expressions
are order-by-order much less dependent on the
renormalization scale $\mu$.  In particular, the full $N^3 L$ summed
expression \tref{S_vec_SN3L} exhibits virtually no dependence on $\mu$, but is seen
to maintain a constant value $S_{RG\Sigma}^{N^3 L} = 1.05372 \pm
0.00004$ over the entire $\sqrt{s}/2 \leq \mu \leq 2\sqrt{s}$ range of
renormalization scale considered.  By contrast the unsummed expression
$S^{N^3 L}$ of \tref{S_vec_N3L} is seen to increase (modestly) over this same range
from 1.0525 to 1.0540.  The point marked FAC in Figure \ref{resum_f8} is the
intersection of the unsummed expressions for $S^{NNL}$ and $S^{N^3 L}$.
This point is the particular choice of $\mu$ at which the
${\cal{O}}(x^3)$ contribution to \tref{S_vec_N3L} vanishes, {\it i.e.}, the point of
fastest apparent convergence (FAC).  It is noteworthy that this FAC
value for $S^{N^3 L}$ is quite close to the RG-summation value
$S_{RG\Sigma}^{N^3 L}$, a result anticipated by Maxwell \cite{maxwell}.
In other words, if one were to use the FAC criterion to reduce the
theoretical uncertainty arising from $\mu$-dependence in the original expression
\tref{S_vec_N3L} for $S^{N^3 L}(\mu)$ over the $\sqrt{s}/2 \leq \mu \leq 2\sqrt{s}$
range considered, the specific value one would obtain [$S_{FAC}^{N^3 L}
= S^{N^3 L} (26 \,{\rm GeV}) = 1.05402$] corresponds very nearly to the RG-summation 
value extracted from \tref{S_vec_SN3L} [$S_{RG\Sigma}^{N^3 L} = 1.05372
\pm 0.00004$].

\begin{figure}[hbt]
\centering
\includegraphics[scale=0.7]{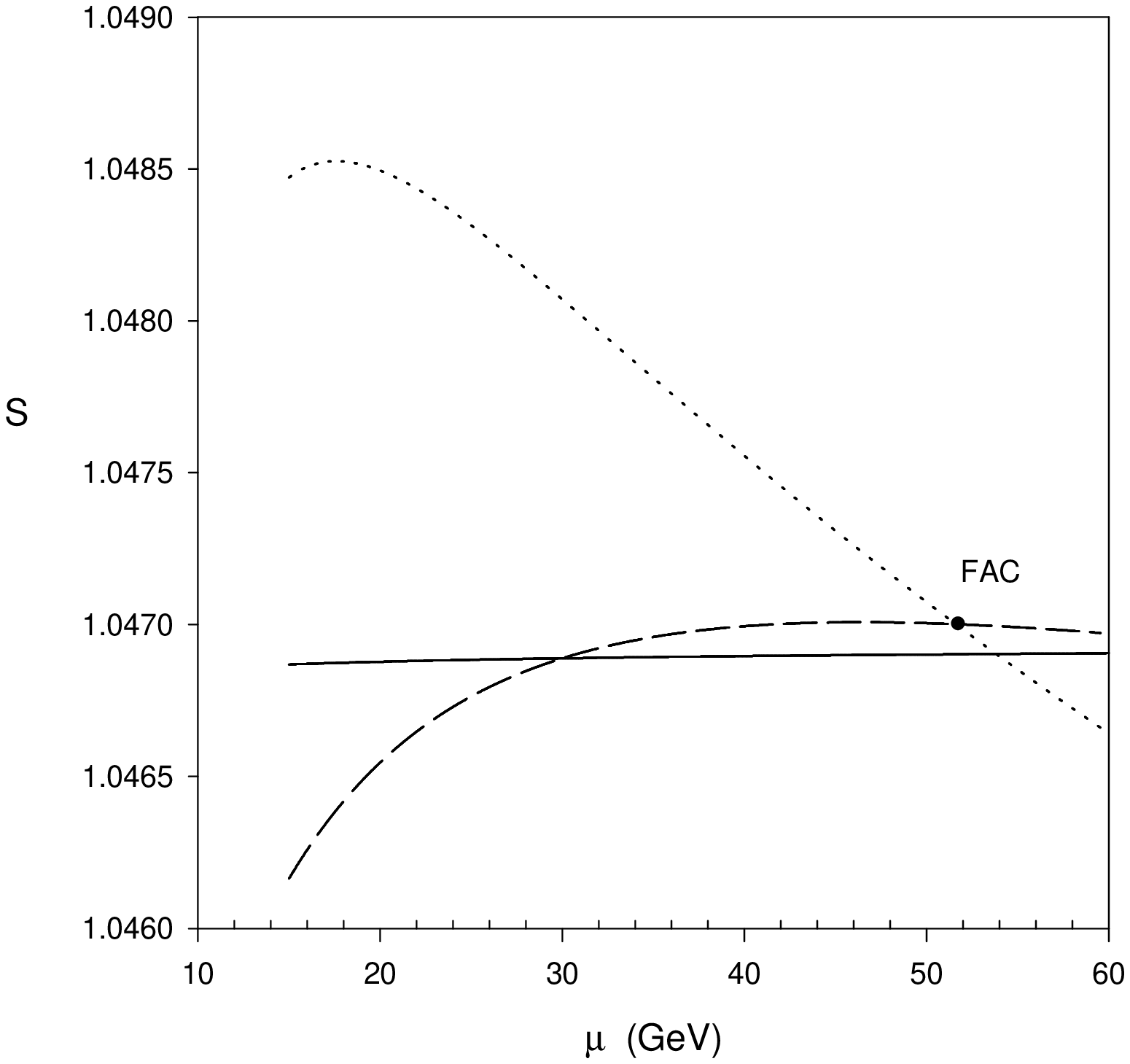}
\caption{
 Comparison of  $S^{NNL}$ (dotted line),  $S^{N^3L}$ (dashed line), and  $S_{RG\Sigma}^{N^3L}$ (solid line)
          expressions for the series $S\left[x(\mu   ), \log\left(\mu^2 /s\right)\right]$ within the  
$n_f = 5$ vector-current
          correlation-function when  $\sqrt{ s} = 30  \,{\rm GeV}$.
}
\label{resum_f10}
\end{figure}

\begin{figure}[hbt]
\centering
\includegraphics[scale=0.7]{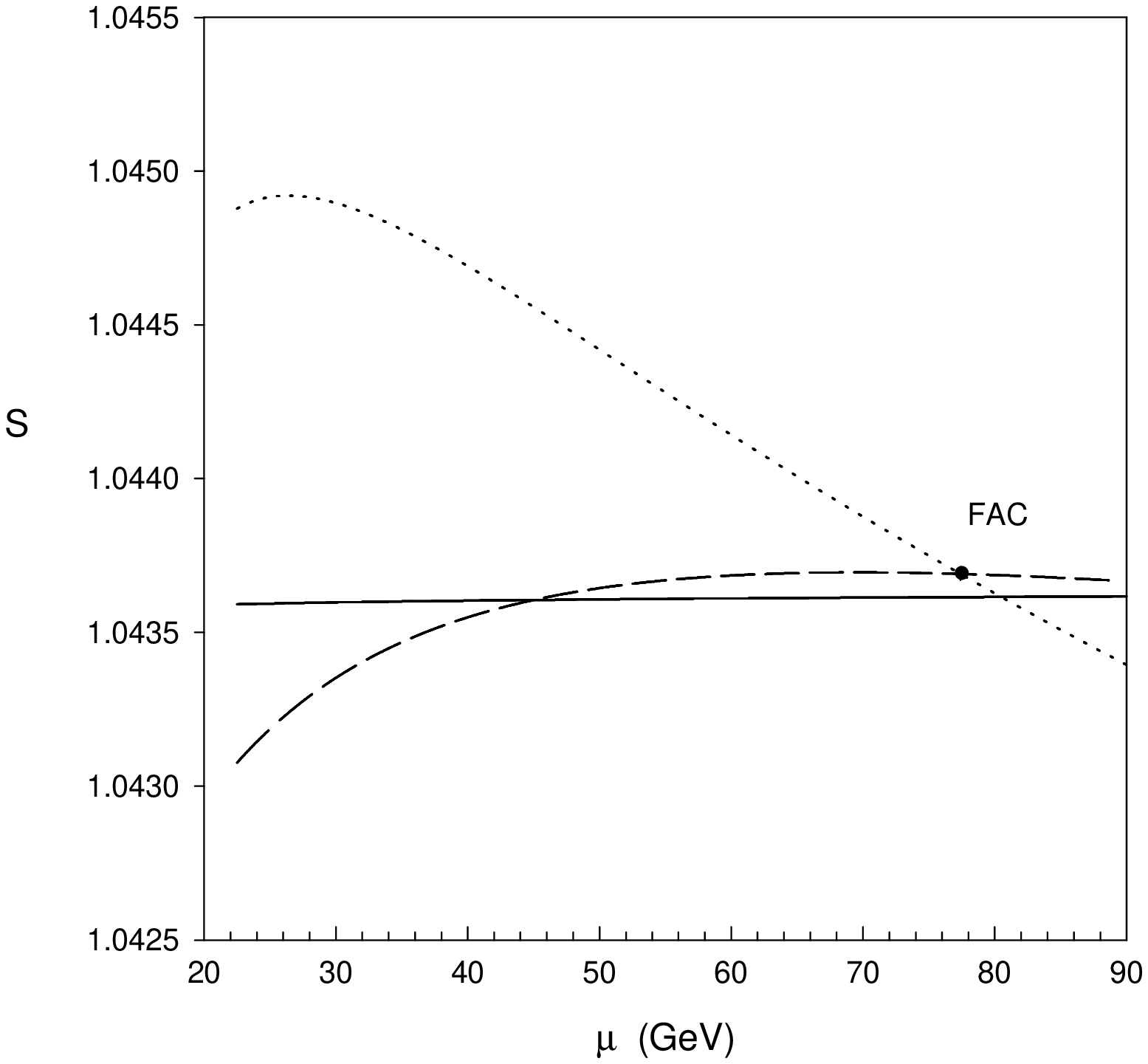}
\caption{
 Comparison of  $S^{NNL}$ (dotted line),  $S^{N^3L}$ (dashed line), and  $S_{RG\Sigma}^{N^3L}$ (solid line)
          expressions for the series $S\left[x(\mu   ), \log\left(\mu^2 /s\right)\right]$ within the  
$n_f = 5$ vector-current
          correlation-function when  $\sqrt{ s} = 45 \,{\rm GeV}$.
}
\label{resum_f11}
\end{figure}

This behaviour is not peculiar to the choice of $s$ in Figs.\ \ref{resum_f8} and \ref{resum_f9}.
In Figures \ref{resum_f10} and \ref{resum_f11}, we consider $S^{N^2 L}$, $S^{N^3 L}$ and
$S_{RG\Sigma}^{N^3 L}$ for $\sqrt{s} = 30 \,{\rm GeV}$ and $\sqrt{s} = 45 \;
\,{\rm GeV}$.  The FAC point each figure occurs at the value of $\mu$ for which
$S^{NNL} (\mu) = S^{N^3 L} (\mu)$, as discussed above.  In both figures
it is evident that $S^{N^3 L}$ has substantially more variation with
$\mu$ than $S_{RG\Sigma}^{N^3 L}$, which is virtually independent of
$\mu$.  Nevertheless, both figures also show that the FAC point of
$S^{N^3 L}$ is very close to the $S_{RG\Sigma}^{N^3 L}$-level value.
For $\sqrt{s} = 30 \,{\rm GeV}$ [Fig.\  \ref{resum_f10}] $S_{FAC}^{N^3 L} = S^{N^3 L}
(\mu=52 \,{\rm GeV}) = 1.04700$ and $S_{RG\Sigma}^{N^3 L} = 1.04689 \pm
0.00002$.  For $\sqrt{s} = 45 \,{\rm GeV}$ [Fig.\  \ref{resum_f11}], $S_{FAC}^{N^3 L} =
S^{N^3 L}(\mu=77 \,{\rm GeV}) = 1.04369$, and $S_{RG\Sigma}^{N^3 L} = 1.04360
\pm 0.00001$.

Similarly, the PMS points in all three figures, corresponding to
maxima of $S^{N^3 L}(\mu)$, are very near the FAC points and also
quite close to the $S_{RG\Sigma}^{N^3 L}$ level.  In Fig.\ \ref{resum_f8} $[\sqrt{s} =
15 \,{\rm GeV}]$, $S_{PMS}^{N^3 L} = S^{N^3 L} (23 \,{\rm GeV}) = 1.05403$; in
Fig.\  \ref{resum_f10} $[\sqrt{s} = 30 \,{\rm GeV}]$, $S_{PMS}^{N^3 L} = S^{N^3 L} (46 \;
\,{\rm GeV}) = 1.04701$; and in Fig.\  \ref{resum_f11} $[\sqrt{s} = 45 \,{\rm GeV}]$, $S_{PMS}^{N^3
L} = S^{N^3 L}(70 \,{\rm GeV}) = 1.04370$.  Of course, equality between $S^{N^3 L}(\mu)$
and $S_{RG\Sigma}^{N^3 L}$ is necessarily exact when $L=0$, {\it i.e.}, when
$\mu^2$ is chosen equal to $s$.  This is indeed the prescription
employed in the standard \cite{pdg} prescription relating $S^{N^3 L}$ to
$R(s)$ \cite{gorishny}:
\begin{equation}%5.10
R(s)=1 + x\left(\sqrt{s}\right) + T_{2,0}~ x^2\!\left(\sqrt{s}\right) + T_{3,0}~ x^3\!\left(\sqrt{s}\right) ,
\end{equation}
where $T_{2,0}$ and $T_{3,0}$ are given in Table \ref{resum_tab1} for $n_f = \{3, 4,
5\}$.  The point here, however, is that this prescription is justified
{\em not} by the $\mu$-invariance of $S^{N^3 L}\left[x(\mu),
\log\left(\mu^2/s\right)\right]$, the truncated perturbative
series, but by that of $S_{RG\Sigma}^{N^3 L}$, the perturbative series
incorporating the closed-form summation of all RG-accessible logarithms
within higher-order terms.  Moreover, the nearly $\mu$-independent result
$S_{RG\Sigma}^{N^3 L}$ appears to be quite close to the result one would
obtain from $S^{N^3 L}\left[ x(\mu), \log \left(\mu^2/s\right)
\right]$ either by imposing FAC or PMS criteria to establish an optimal
value of $\mu$, as graphically evident from Figs.\ \ref{resum_f10} and \ref{resum_f11}.  That
optimal value for $\mu$, however, is {\em not} $\mu=\sqrt{s}$, but a
substantially larger value of $\mu$ for each case considered.

\begin{figure}[hbt]
\centering
\includegraphics[scale=0.7]{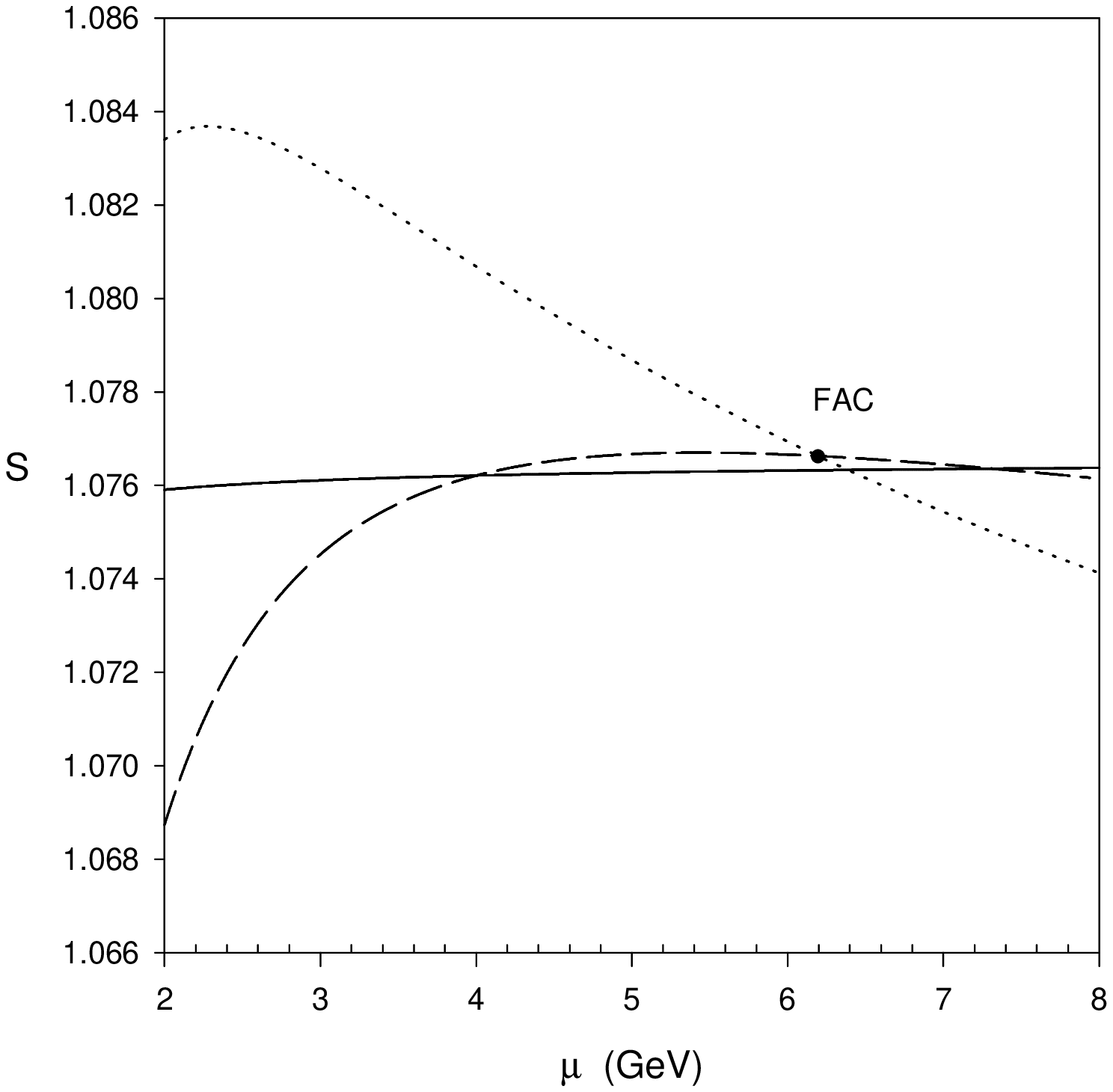}
\caption{
 Comparison of  $S^{NNL}$ (dotted line),  $S^{N^3L}$ (dashed line), and  $S_{RG\Sigma}^{N^3L}$ (solid line)
          expressions for the series $S\left[x(\mu   ), \log\left(\mu^2 /s\right)\right]$ within the  
$n_f = 4$ vector-current
          correlation-function when  $\sqrt{ s} = 4 \,{\rm GeV}$.
}
\label{resum_f12}
\end{figure}

\begin{figure}[hbt]
\centering
\includegraphics[scale=0.7]{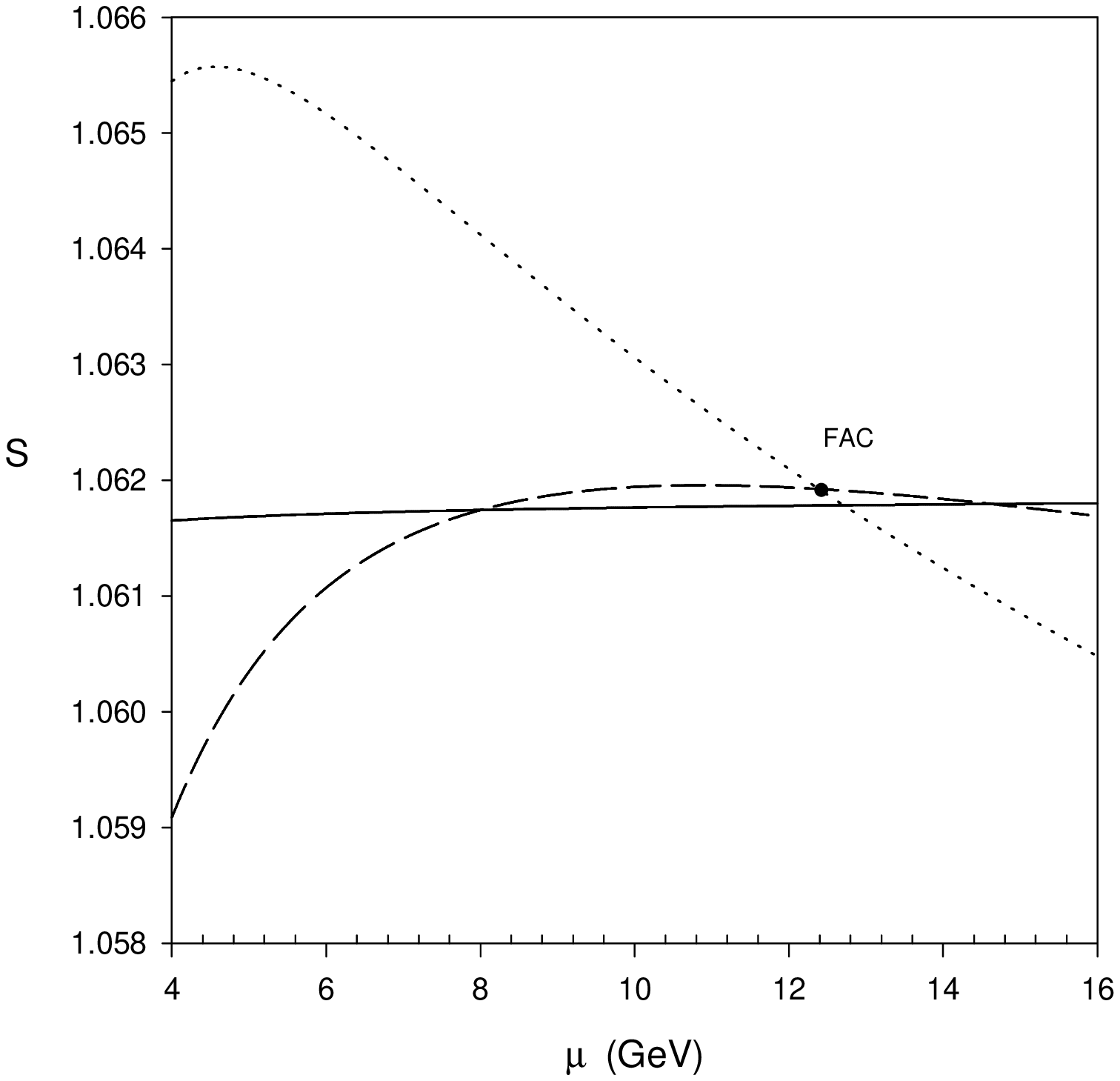}
\caption{
 Comparison of  $S^{NNL}$ (dotted line),  $S^{N^3L}$ (dashed line), and  $S_{RG\Sigma}^{N^3L}$ (solid line)
          expressions for the series $S\left[x(\mu   ), \log\left(\mu^2 /s\right)\right]$ within the  
$n_f = 4$ vector-current
          correlation-function when  $\sqrt{ s} = 8 \,{\rm GeV}$.
}
\label{resum_f13}
\end{figure}

These results are not peculiar to the choice $n_f = 5$.  Using the Table \ref{resum_tab1}
 entries for the parameters $T_{2,0}, \; T_{3,0}, \; G, P, U$ and $V$
appearing in \tref{S2_pole} and \tref{S3_pole}, we find for $n_f = 3$ that
\begin{equation}%5.11
\begin{split}
S_{RG\Sigma}^{N^3 L} [x, L]  =  1 &+ \frac{x}{1-\frac{9}{4} x L}
 +  \frac{x^2\left[1.63982 - \frac{16}{9} \log \left( 1-\frac{9}{4} x
L \right) \right] + x^3 \left(-\frac{3397}{2592}\right)}{\left( 1 -
\frac{9}{4} x L\right)^2}
\\
 &+  \frac{x^3 \left[-8.97333 - 8.99096 \log \left( 1-\frac{9}{4} x L
\right) + \frac{256}{81} \log^2 \left( 1 - \frac{9}{4} x L \right)
\right]}{\left( 1 - \frac{9}{4} x L \right)^3}
%\nonumber\\
\end{split}
\end{equation}
and for $n_f = 4$ that
\begin{equation}%5.12
\begin{split}
S_{RG\Sigma}^{N^3 L} [x, L]  =  1 &+ \frac{x}{1 -\frac{25}{12} x
L}
 +  \frac{x^2 \left[ 1.52453 - \frac{77}{50} \log \left( 1 -
\frac{25}{12} x L \right) \right] + x^3 \left(- \frac{121687}{180000}
\right)}{\left( 1 - \frac{25}{12} x L \right)^2}
\\
& +  \frac{x^3 \left[ -11.0096 - 7.06715 \log \left( 1 - \frac{25}{12} x L
\right) + \frac{5929}{2500} \log^2 \left( 1 - \frac{25}{12} x L \right)
\right]}{\left( 1 - \frac{25}{12} x L \right)^3}
\end{split}
\tlabel{S_4f_SN3L}
\end{equation}
In Figures \ref{resum_f12} and \ref{resum_f13} we display the expression \tref{S_4f_SN3L} for
$S_{RG\Sigma}^{N^3 L}$ together with (unsummed $n_f = 4$ expressions
for)
$S^{N^3 L}$ and $S^{NNL}$ for $\sqrt{s} = 4 \,{\rm GeV}$ and $\sqrt{s} = 8
\,{\rm GeV}$.  
To generate these figures, we evolve $x(\mu)$ using an initial condition
$x(4.17\,{\rm GeV})=0.0716218$ \cite{ahmady}
appropriate for $n_f=4$ and obtained from the threshold matching conditions
\cite{cks} to the $n_f=5$ running couplant $x(4.17\,{\rm GeV})=0.0715492$ evolved from
$x\left(M_Z\right)=0.118/\pi$ \cite{pdg}.
As in Figs.\ \ref{resum_f10} and \ref{resum_f11} we see that $S_{PMS}^{N^3 L}$ and
$S_{FAC}^{N^3 L}$ are both close to the very-nearly constant
$S_{RG\Sigma}^{N^3 L}$ level, although the PMS and FAC values for $\mu$
are substantially larger than $\sqrt{s}$ in each case.  

To conclude, the primary result of interest is that closed-form
summation of all RG-accessible logarithms to any given order of
perturbation theory leads to expressions ({\it e.g.} Fig.\ \ref{resum_f9}) that 
order-by-order are substantially less scale-dependent than the corresponding
truncated series (Fig.\ \ref{resum_f8}).  The scale-independence of $S_{RG\Sigma}^{N^3
L}$ supports the prescription of choosing $\mu = \sqrt{s}$ in the
unsummed series $S^{N^3 L}\left[ \mu^2, \; \log (\mu^2 / s)\right]$,
since the summed and unsummed series coincide at this value of $\mu$.
However, the unsummed series $S^{N^3 L}$ still exhibits noticeable scale
dependence.  The use of FAC and PMS criteria to find an optimal value
for $\mu$ for $S^{N^3 L}$ leads to values for this unsummed series which are quite close
to its RG-summation value.

\section{Other Perturbative Applications}\label{pert_sec}

\renewcommand{\theequation}
{6.\arabic{equation}}
\setcounter{equation}{0}

\subsection{Momentum Space QCD Static Potential}
The momentum space expression for the perturbative portion of the QCD
static potential function
\begin{equation}%6.1
V(r) = \int \frac{\mathrm{d}^3 q}{(2\pi)^3}\, \mathrm{e}^{\mathrm{i} \vec{q} \cdot \vec{r}} \left( -
\frac{16\pi^2}{3\vec{q}^{\; 2}}\right) W \left[ \frac{\alpha_s(\mu)}{\pi}, \log
\left( \frac{\mu^2}{\vec{q}^{\; 2}} \right) \right]
\end{equation}
is given by the integrand series
\begin{equation}%6.2
W[x, L]  =  x + \left( T_{2,0} + T_{2,1} L \right) x^2 
 +  \left( T_{3,0} + T_{3,1} L + T_{3,2} L^2 \right) x^3 + \ldots
\tlabel{W_series}
\end{equation}
where $x \equiv \alpha_s (\mu) / \pi$, $L = \log\left(\mu^2 / \vec{q}^{\,2}\right)$, and 
where the series coefficients within \tref{W_series} are
\cite{schroder}
\begin{equation}
\begin{split}%6.3
T_{2,0} &= 31/12 - 5 n_f / 18, \; \; T_{2,1} = \beta_0 = 11/4 - n_f / 6,
\\
T_{3,0} &= 28.5468 - 4.14714 n_f + 25 n_f^2 / 324,\;
T_{3,1} = 247/12 - 229 n_f / 72 + 5 n_f^2 / 54, \; \; T_{3,2} =
\beta_0^2.
\end{split}
\end{equation}
The function $W[x, L]$ is shown in \cite{static} to satisfy the same RG
equation \tref{RG_vec} as the semileptonic $b \rightarrow u$ and $b \rightarrow
c$ decay rate in the pole mass scheme.  Consequently, the closed form
summation of all RG-accessible logs is given by eqs.\ \tref{S1_pole}--\tref{S3_pole}, with the
constants $G, P, U$ and $V$ as given by eqs.\ \tref{G_pole_res}--\tref{V_pole_res}.  Since $T_{0,0}$,
as defined by the generic series form \tref{pert_series}, is zero for the series
\tref{W_series}, the series $S_0 = \sum_{n=0}^\infty T_{n,n} (x L)^n$ is trivially
zero.  Thus, the RG-summation of $W[x, L]$ is
\begin{equation}%6.4
W_{RG\Sigma} = x S_1 [x L] + x^2 S_2 [x L] + x^3 S_3 [x L].
\end{equation}
Noting that $T_{1,0} = 1$ in \tref{W_series}, we find for $n_f = \{3, 4, 5\}$
that 
\begin{gather}%6.5
\begin{split}
W_{RG\Sigma}^{n_f=5}  = & \frac{x}{(1-\frac{23}{12} xL)}
 +  \frac{x^2 \left[ \frac{43}{36} - \frac{29}{23} \log \left( 1-
\frac{23}{12} xL \right) \right] + x^3 \left[
\frac{17521}{152352}\right]}{\left( 1 - \frac{23}{12} xL\right)^2}
\\
& +  \frac{x^3 \left[9.62511 - \frac{43819}{9522} \log \left( 1-
\frac{23}{12} x L\right) + \frac{841}{529} \log^2 \left( 1 -
\frac{23}{12} xL \right) \right]}{\left( 1-\frac{23}{12} xL
\right)^3}
\end{split}
\tlabel{W_3f}
\\[5pt]
\begin{split}%6.6
W_{RG\Sigma}^{n_f=4}  = & \frac{x}{(1-\frac{25}{12} xL)}
 +  \frac{x^2 \left[ \frac{53}{36} - \frac{77}{50} \log \left( 1-
\frac{25}{12} xL \right) \right] - x^3 \left[
\frac{121687}{180000}\right]}{\left( 1 - \frac{25}{12} xL\right)^2}
\\
& +  \frac{x^3 \left[13.8688 - \frac{77693}{11250} \log \left( 1-
\frac{25}{12} x L\right) + \frac{5929}{2500} \log^2 \left( 1 -
\frac{25}{12} xL \right) \right]}{\left( 1-\frac{25}{12} xL
\right)^3}
\end{split}
\\[5pt]
\begin{split}%6.7
W_{RG\Sigma}^{n_f=3}  = & \frac{x}{\left(1-\frac{9}{4} xL\right)}
 +  \frac{x^2 \left[ \frac{7}{4} - \frac{16}{9} \log \left( 1-
\frac{9}{4} xL \right) \right] - x^3 \left[
\frac{3397}{2592}\right]}{\left( 1 - \frac{9}{4} xL\right)^2}
\\
& +  \frac{x^3 \left[18.1104 - \frac{760}{81} \log \left( 1-
\frac{9}{4} x L\right) + \frac{256}{81} \log^2 \left( 1 -
\frac{9}{4} xL \right) \right]}{\left( 1-\frac{9}{4} xL
\right)^3}
\end{split}
\end{gather}

\begin{figure}[hbt]
\centering
\includegraphics[scale=0.7]{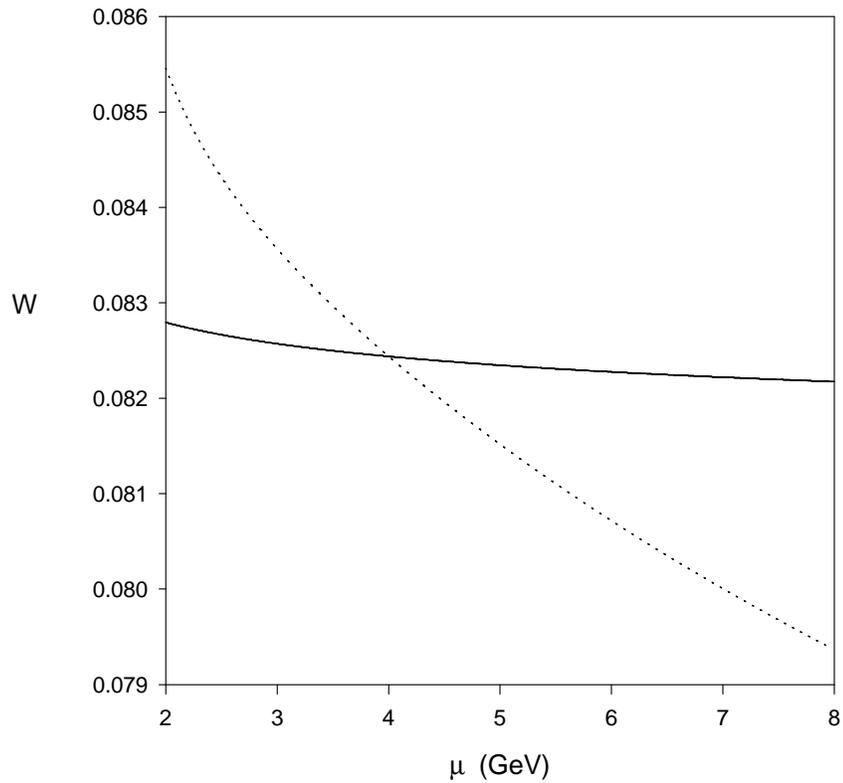}
\caption{
 The momentum-space $n_f = 5$ static-potential function series $W^{NNL}$ (dotted line),
          and the corresponding RG-summation $W^{NNL}_{RG\Sigma}$   (solid line) with 
  $\left\vert\vec q\right\vert =4 \,{\rm GeV}$. 
}
\label{resum_f14}
\end{figure}

In Fig.\ \ref{resum_f14} we plot both the truncated series $W^{NNL} [x, L]$,
consisting of the terms explicitly appearing in \tref{W_series}, as well as the
corresponding RG-summed series \tref{W_3f} for the $n_f = 5$ case with
$|\vec{q}| = 4 \,{\rm GeV}$.  As before, we determine $x(\mu)$ from eq.\ \tref{beta_def}
with the initial condition $\alpha_s (4.17 \,{\rm GeV})/\pi = 0.071549$
devolving from $\alpha_s (M_z) = 0.118000$.  Although the truncated
series varies only modestly [$0.0856 \geq W^{NNL} \geq 0.0795$] over the
range $|\vec{q}| / 2 < \mu < 2 | \vec{q} |$, the RG-summed series is
seen to exhibit less than 10\% of the unsummed series' variation over
this same range of $\mu$ $[0.08295 \geq W_{RG\Sigma} \geq 0.08233]$.

\subsection{Gluonic Scalar Correlation Function}\label{scalar_glue_sec}
The imaginary part of the correlator for gluonic scalar currents
\begin{equation}%6.8
j_G(y) = \frac{\beta(x(\mu))}{\pi x(\mu)\beta_0} G_{\mu\nu}^a (y) G^{\mu\nu, a} (y)
\end{equation}
enters QCD sum rules pertinent to scalar glueball properties \cite{glue}, and is given by
$(s \equiv p^2, \; x(\mu) = \alpha_s (\mu)/\pi)$
\begin{equation}%6.9
{\rm Im}\Pi_G (s)  =  {\rm Im}\left\{ \mathrm{i} \int \mathrm{d}^4 y \; \mathrm{e}^{\mathrm{i} p\cdot y} <0|T j_G (y)
j_G (0)|0>\right\}
 =  \frac{2 x^2(\mu) s^2}{\pi^3} S \left[x(\mu), \; \log\left(\mu^2 / s\right)
\right]
\tlabel{S_G_def}
\end{equation}
where
\begin{equation}%6.10
S[x, L] = 1 + \sum_{n=1}^\infty \sum_{m=0}^n T_{n,m} x^n L^m .
\tlabel{S_G_series}
\end{equation}
The leading coefficients $T_{n,m}$ within \tref{S_G_series} can be extracted from a
three-loop calculation by Chetyrkin, Kniehl and Steinhauser
\cite{kniehl, ces} and are tabulated in Table \ref{resum_tab2}.

\begin{figure}[hbt]
\centering
\includegraphics[scale=0.7]{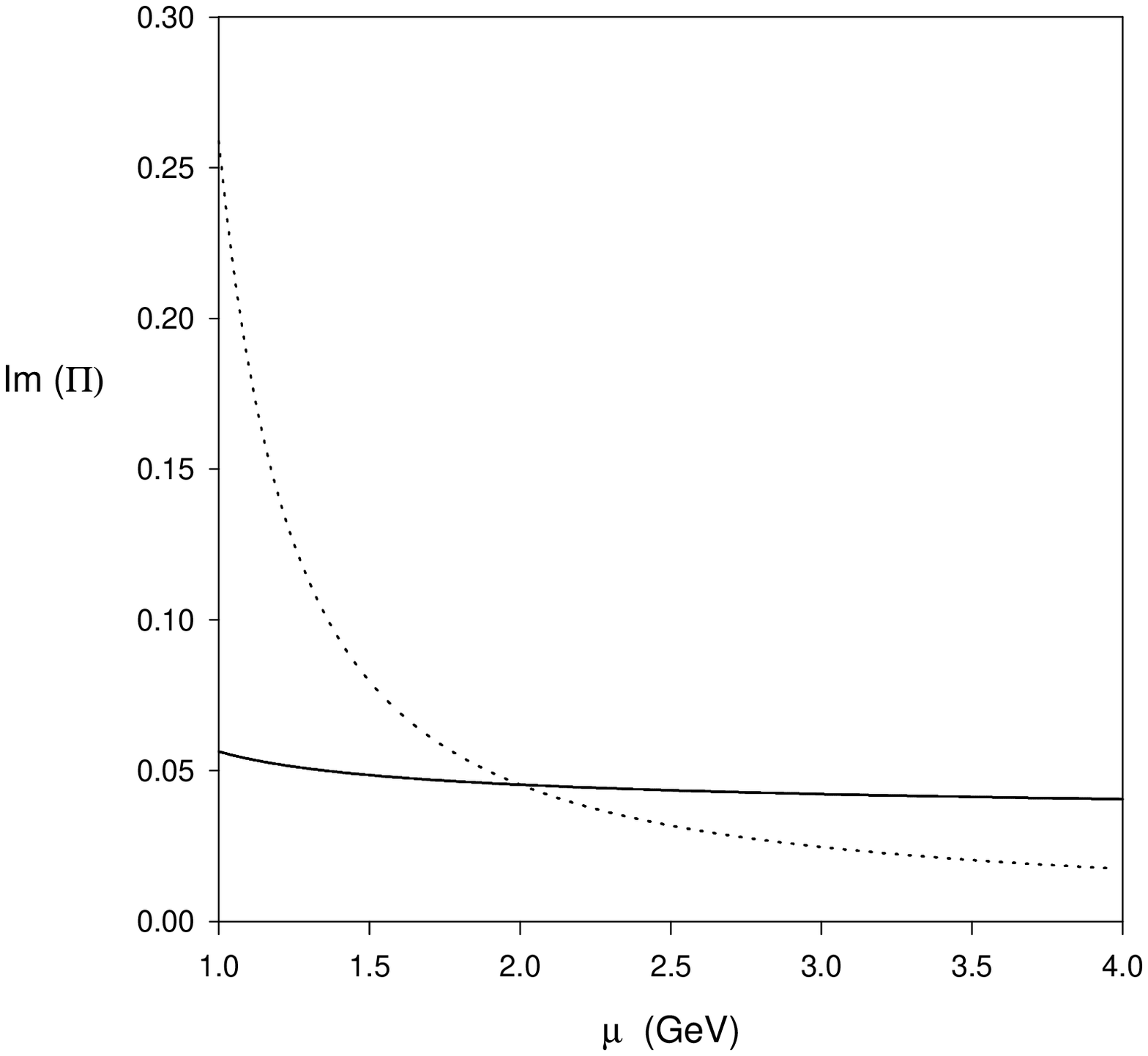}
\caption{
 The imaginary part of the gluonic scalar-current correlation function (\protect\ref{S_G_def}), as
          obtained from $x^2(\mu)\,S^{NNL}$ (dotted line) and from  $x^2(\mu)\,S^{NNL}_{RG\Sigma}$  
(solid line) 
with $n_f = 3$
          and  $\sqrt{s} = 2\,{\rm GeV}$.
}
\label{resum_f15}
\end{figure}

The correlator $Im\Pi_G (s)$ is RG-invariant:  $\mu^2 \mathrm{d} Im\Pi_G (s) / \mathrm{d}\mu^2
= 0$.  Consequently the series \tref{S_G_series} can be shown to satisfy the RG-
equation
\begin{equation}%6.11
\left( \frac{\partial}{\partial L} + \beta(x) \frac{\partial}{\partial
x} + \frac{2\beta(x)}{x} \right) S[x, L] = 0.
\tlabel{glue_RG}
\end{equation}
Upon substituting \tref{S_G_series} into \tref{glue_RG}, it is straightforward to show that
the aggregate coefficients of $x^n L^{n-1},\; x^n L^{n-2}$ and $x^n
L^{n-3}$ respectively vanish provided
\begin{gather}%6.12
n T_{n,n} - \beta_0 (n+1) T_{n-1, \; n-1} = 0,
\tlabel{glue_RG_e1}
\\
(n-1) T_{n, n-1} - \beta_0 (n+1) T_{n-1, \; n-2} - \beta_1 n T_{n-2, \;
n-2} = 0,
\tlabel{glue_RG_e2}
\\
(n-2) T_{n, n-2} - \beta_0 (n+1) T_{n-1, \; n-3} - \beta_1 n T_{n-2, \; n-
3} - \beta_2 (n-1) T_{n-3, \; n-3} = 0.
\tlabel{glue_RG_e3}
\end{gather}
We employ the definitions \tref{S0_decay_def}, \tref{S1_decay_def} and \tref{S2_decay_def} for $S_0$, $S_1$ and
$S_2$.  By multiplying \tref{glue_RG_e1} by $u^{n-1}$, \tref{glue_RG_e2} by $u^{n-2}$ and \tref{glue_RG_e3}
by $u^{n-3}$ and summing from $n=1, 2$ and $3$ respectively, we obtain
the following three linear differential equations
\begin{gather}%6.15
(1-\beta_0 u) \frac{\mathrm{d} S_0}{\mathrm{d}u} - 2 \beta_0 S_0 = 0
\\
(1-\beta_0 u) \frac{\mathrm{d} S_1}{\mathrm{d}u} - 3 \beta_0 S_1 = \beta_1 \left[
u\frac{\mathrm{d} S_0}{\mathrm{d}u} + 2 S_0 \right],
\\
(1-\beta_0 u) \frac{\mathrm{d} S_2}{\mathrm{d}u} - 4 \beta_0 S_2 = \beta_1 \left[
u\frac{\mathrm{d} S_1}{\mathrm{d}u} + 3 S_1 \right] 
+ \beta_2 \left[ u \frac{\mathrm{d} S_0}{\mathrm{d}u} + 2 S_0 \right].
\end{gather}
Given the $u=0$ initial conditions $S_0 = 1, \; S_1 = T_{1,0}, \; S_2 =
T_{2,0}$ and setting $u= x L$, we obtain the following RG-summed version
of the series \tref{S_G_series} to NNL order:
\begin{equation}%6.18
\begin{split}
S_{RG\Sigma}^{NNL}  = & S_0 [x L] + x S_1 [x L] + x^2 S_2 [x L]
\\
 = & \frac{1}{(1-\beta_0 xL)^2} + \frac{x\left[ T_{1,0} -
\frac{2\beta_1}{\beta_0} \log (1-\beta_0 x L) \right]+ x^2 \left[ \frac{2\beta_1^2}{\beta_0^2} - \frac{2\beta_2}{\beta_0}\right]}{(1-\beta_0 x
L)^3}
\\
& +  x^2 \frac{\left[ \left( T_{2,0} - \frac{2\beta_1^2}{\beta_0^2} +
\frac{2\beta_2}{\beta_0} \right) - \left( \frac{3 T_{1,0}
\beta_1}{\beta_0} + \frac{2\beta_1^2}{\beta_0^2} \right) \log (1-\beta_0
x L)
 +   (3\beta_1^2 / \beta_0^2) \log^2 (1-\beta_0 x L) \right]}{
(1-\beta_0 x L)^4}
\end{split}
\tlabel{S_glue_SNNL}
\end{equation}
where $x = \alpha_s (\mu) / \pi$ and $L = \log (\mu^2 / s)$, as before,
and where the $\beta_k$ are as defined in \tref{beta_def}.  In Figure \ref{resum_f15} we compare
the $\mu$-dependence of $x^2 (\mu) S_{RG\Sigma}^{NNL}$ to that of the
corresponding truncated series
\begin{equation}%6.19
x^2 (\mu) S^{NNL} [x, L] = x^2 \left[1+\left(T_{1,0} + T_{1,1} L\right)x + 
\left(T_{2,0} + T_{2,1} L + T_{2,2} L^2\right) x^2\right]
\tlabel{S_glue_NNL}
\end{equation} 
for the $n_f = 3$ case with $\sqrt{s} = 2 \,{\rm GeV}$.  The evolution of
$x(\mu)$ is assumed to follow an $n_f = 3$ $\beta$-function with the
initial condition $\alpha_s (m_\tau) / \pi = 0.33/\pi$ \cite{alpha}.  As
evident from the figure, the severe $\mu$-dependence of $x^2 S^{NNL}$ is
considerably diminished by RG-summation.  The RG-summed expression [eq.\
\tref{S_glue_SNNL} multiplied by $x^2(\mu)$] falls from 0.056 to 0.041 as $\mu$
increases from $1 \,{\rm GeV}$ to $4 \,{\rm GeV}$.  By contrast, the unsummed
expression \tref{S_glue_NNL} falls precipitously from 0.259 to 0.017, a factor of
15, over the same range of $\mu$.  This unphysical dependence on
renormalization scale suggests that $S^{NNL}$ [as in \tref{S_glue_NNL}] be replaced
by \tref{S_glue_SNNL} within sum rule approaches to the lowest-lying scalar gluonium state.

\subsection{Cross-Section $\sigma(W_L^+ W_L^- \rightarrow Z_L Z_L)$}
The scattering of two longitudinal $W$'s into two longitudinal $Z$'s is
mediated by the Higgs particle of standard-model electroweak physics.
Assuming a single Higgs particle (devolving from the single doublet
responsible for electroweak symmetry breaking), one finds the 
cross-section for this process at very high energies $(s >> M_H^2)$ to be
\begin{equation}%6.20
\sigma(s) = \frac{8\pi^2}{9s} g^2 (\mu) S[g(\mu), \log\left(\mu^2 /s\right)],
\end{equation}
where $g(\mu) = 6 \lambda_{\overline{\rm{MS}}} (\mu) / 16\pi^2$, the quartic
scalar couplant of the single-doublet 
standard model,\footnote{$\lambda_{\overline{\rm{MS}}} (\mu)$ is perturbatively related to
its on-mass-shell value $G_F M_H^2 / \sqrt{2}$, as discussed in
\cite{nierste}.} and where
the series $S$ is \cite{nierste}
\begin{equation}%6.21
S[g, L] = 1 + \sum_{n=1}^\infty \sum_{m=0}^n T_{m, n} g^n L^m
\end{equation}
with $g=g(\mu)$,and $ L= \log(\mu^2 / s)$. The constants $T_{n, m}$
are fully known to ${\rm NNL}$ order \cite{nierste} 
\begin{equation}%6.22
T_{1,0} = -10.0, \; T_{1,1} = -4, \; T_{2,0} = 93.553 + \frac{2}{3} 
\log\left(s/M_H^2\right), \; T_{2,1} = 68.667, \; T_{2,2} = 12.
\end{equation}
The RG-invariance of the physical cross-section $\sigma$ implies that
the series $S$ satisfies the RG equation
\begin{equation}%6.23
\left( \frac{\partial}{\partial L} + \beta(g) \frac{\partial}{\partial
g} + \frac{2\beta(g)}{g} \right) S[g, L] = 0
\end{equation}
This is the same RG-equation as \tref{glue_RG} characterizing the gluonic scalar
correlator, but with the $\beta$-function appropriate for the 
single-Higgs-doublet $\overline{\rm{MS}}$ quartic scalar couplant $g(\mu)$ \cite{kleinert}:
\begin{equation}%6.24
\begin{split}
\mu^2 \frac{\mathrm{d}g}{\mathrm{d}\mu^2} &= \beta(g) \equiv -\beta_0 g^2 - \beta_1 g^3 -
\beta_2 g^4 \ldots,
\\
\beta_0 &= -2, \; \beta_1 = \frac{13}{3}, \; \beta_2 = -27.803
\end{split}
\end{equation}
Consequently the RG-summed version of $S$ is given by \tref{S_glue_SNNL}:
\begin{equation}%6.25
\begin{split}
S_{RG\Sigma}  = & \frac{1}{(1+2gL)^2} + \frac{g \left[-10+\frac{13}{3} \log (1+2gL)\right] - 18.4141g^2}{(1+2 gL)^3}
\\
& +  \frac{g^2 \left[111.967+\frac{2}{3}\log\left(\frac{s}{M_H^2}\right) - \frac{1339}{18} \log (1+2gL) + \frac{169}{12} \log^2 (1+2gL)\right]}{(1+2gL)^4}
\end{split}
\end{equation} 

\subsection{Higgs Decay $H \rightarrow gg$}
Higher-order expressions for the decay of a Higgs boson into two gluons
have been obtained and studied both outside \cite{cpviol} and within
\cite{higgs, higgsier, higgsiest} the context of a standard-model 
single-doublet Higgs field.  In the limit $M_H^2 << 4M_t^2, \; \; M_b = 0$ the latter decay rate
is of the form
\begin{equation}%6.26
\Gamma = \left[ \sqrt{2} G_F M_H^3 / 72\pi \right] x^2 (\mu) 
S\left[ x(\mu), \log \left( \mu^2 / m_t^2 (\mu) \right)\!, \log\left( M_H^2 / M_t^2\right) \right].
\tlabel{Gam_higgs}
\end{equation}
Capitalized masses $\left(M_H, M_t\right)$ denote RG-invariant pole masses, whereas $m_t(\mu)$ is the running t-quark mass.  The series
$S$ within \tref{Gam_higgs} is of the generic form \tref{pert_series} with $L\equiv \log
\left(\mu^2 / m_t^2 (\mu) \right)$, but the coefficients $T_{n,m}$ are
now dependent upon the RG-invariant logarithm $T \equiv \log \left(
M_H^2 / M_t^2\right)$.  Using six active flavours to accommodate the
running of $m_t (\mu)$, one can extract from ref.\ \cite{higgs} the following
two subleading orders of series coefficients within $S$ \cite{higgsier}:
\begin{equation}%6.27
\begin{split}
T_{0,0} &= 1, \; T_{1,0} (T) = \frac{215}{12} - \frac{23}{6} T, \;
T_{1,1} = \frac{7}{2}, 
\\
T_{2,0} (T) &= 146.8912 - \frac{4903}{48} T + \frac{529}{48} T^2,
T_{2,1} (T) = \frac{1445}{16} - \frac{161}{8} T, \; T_{2,2} =
\frac{147}{16}
\end{split}
\tlabel{higgs_coeffs}
\end{equation}
RG-invariance of the physical decay rate $\left( \mu^2 d\Gamma / d\mu^2
= 0\right)$ implies the following RG-equation for the series $S[x, L,
T]$ within \tref{Gam_higgs}:
\begin{equation}%6.28
\left[ \left( 1 - 2\gamma_m (x) \right) \frac{\partial}{\partial L} +
\beta (x) \frac{\partial}{\partial x} + \frac{2\beta(x)}{x} \right] S [x,
L, T] = 0.
\tlabel{higgs_RG}
\end{equation}
The $n_f = 6$ values for the $\overline{{\rm MS}}$ $\beta$- and $\gamma_m$-functions are
\begin{equation}%6.29
\begin{split}
\beta(x) &= -\left( \beta_0 x^2 + \beta_1 x^3 + \beta_2 x^4+\ldots\right), \; \;
\beta_0 = 7/4, \beta_1 = 13/8, \beta_2 = -65/128
\\
\gamma_m (x) &= - \left( \gamma_0 x + \gamma_1 x^2 + \gamma_2 x^3 +\ldots\right),
\; \; \gamma_0 = 1, \gamma_1 = 27/8, \gamma_2 = 4.83866.
\end{split}
\tlabel{beta_6f}
\end{equation}
Upon substituting the series $S$, as described above, into \tref{higgs_RG}, one
finds that the net coefficients of $x^n L^{n-1}, x^n L^{n-2}$ and $x^n L^{n-3}$ 
on the left-hand side of \tref{higgs_RG} vanish provided the following recursion
relations are respectively upheld:
\begin{gather}%6.30
n T_{n, n} - \beta_0 (n+1) T_{n-1, n-1} = 0
\tlabel{higgs_RG_e1}
\\
(n-1) T_{n, n-1} + 2\gamma_0 (n-1) T_{n-1, n-1} - \beta_0 (n+1) T_{n-1,
n-2} - \beta_1 n T_{n-2, n-2} = 0
\tlabel{higgs_RG_e2}
\\
\begin{split}
0=&(n-2) T_{n, n-2} + 2\gamma_0 (n-2) T_{n-1, n-2} + 2\gamma_1 (n-2) T_{n-
2, n-2} \\
&-\beta_0 (n+1) T_{n-1, n-3} - \beta_1 n T_{n-2, n-3} - \beta_2 (n-1)
T_{n-3, n-3} \quad.
\end{split}
\tlabel{higgs_RG_e3}
\end{gather}
By multiplying \tref{higgs_RG_e1} by $u^{n-1}$, \tref{higgs_RG_e2} by $u^{n-2}$ and \tref{higgs_RG_e3} by
$u^{n-3}$, and by summing each equation from $n=1, 2$ and $3$,
respectively, we obtain the following linear differential equations for
the summations $S_0$ \tref{S0_decay_def}, $S_1$ \tref{S1_decay_def} and $S_2$ \tref{S2_decay_def}:
\begin{gather}%6.33
(1-\beta_0 u) \frac{\mathrm{d} S_0}{\mathrm{d}u} - 2\beta_0 S_0 = 0
\\
(1-\beta_0 u) \frac{\mathrm{d} S_1}{\mathrm{d}u} - 3\beta_0 S_1 = \beta_1 \left( u 
\frac{\mathrm{d}S_0}{\mathrm{d}u} + 2 S_0 \right) - 2\gamma_0 \frac{\mathrm{d} S_0}{\mathrm{d}u}
\\
(1-\beta_0 u) \frac{\mathrm{d} S_2}{\mathrm{d}u} - 4\beta_0 S_2  =  \beta_1 
\left( u\frac{\mathrm{d}S_1}{\mathrm{d}u} + 3 S_1 \right) - 2\gamma_0 
\frac{\mathrm{d} S_1}{\mathrm{d}u}
 +  \beta_2 \left( u \frac{\mathrm{d} S_0}{\mathrm{d}u} + 2 S_0\right) - 2\gamma_1
\frac{\mathrm{d} S_0}{\mathrm{d}u}
\end{gather}
Given the $u=0$ initial conditions $S_0 = 1$, $S_1 = T_{1,0} (T)$, $S_2
= T_{2,0} (T)$, one can solve for $S_0 [u]$, $S_1 [u]$ and $S_2 [u]$.  As
before, all-orders summation of the RG-accessible logarithms within the
series $S[x, L, T]$ is now possible, given the explicit form of $T_{1,0}
(T)$ and $T_{2,0} (T)$ in \tref{higgs_coeffs} and the explicit $\beta$- and 
$\gamma$-function coefficients in \tref{beta_6f}.  We thus find that
\begin{equation}%6.36
\begin{split}
S_{RG\Sigma}^{NNL}  = & S_0 [x L] + x S_1 [x L] + x^2 S_2 [x L]
\\[5pt]
 = & \frac{1}{\left( 1 - \frac{7}{4} xL\right)^2} + \frac{x\left[
\frac{215}{12} - \frac{23}{6} T + \frac{15}{7} \log \left( 1 -
\frac{7}{4} xL \right) \right] + \frac{9479}{784} x^2}{\left( 1 -
\frac{7}{4} x L\right)^3}
\\[5pt]
& +  \frac{x^2 \left[ 134.801 - \frac{4903}{48} T + \frac{529}{48} T^2
+ \left( \frac{21675}{392}-\frac{345}{28} T \right) \log \left( 1 -
\frac{7}{4} xL \right) + \frac{675}{196} \log^2 \left( 1 - \frac{7}{4}
xL \right) \right]}{\left( 1 - \frac{7}{4} xL\right)^4}
\end{split}
\end{equation}

\subsection{ Fermionic Scalar Correlation Function}
The imaginary part of the RG-invariant correlator for the fermionic
scalar current
\begin{equation}%6.37
j_s (y) = m\overline{\Psi} (y) \Psi (y)
\end{equation}
is
\begin{equation}%6.38
{\rm Im}\; \Pi (s)  =  {\rm Im}\left[ \mathrm{i} \int \mathrm{d}^4 y \;  
\mathrm{e}^{\mathrm{i} p\cdot y} <0|T j_s (y) j_s
(0)|0>\right]
 =  \frac{3s}{8\pi} m^2(\mu) S \left[ x (\mu), \; \log\left(\mu^2 / s\right)
\right],
\tlabel{scalar_quark_corr}
\end{equation}
where the series $S [x, L]$ is of the form \tref{pert_series} and has been fully
calculated to $N^3 L$ order \cite{scalar}.  For $n_f = \{3, 4, 5\}$ the
series coefficients $T_{n,m}$ are tabulated for $(n, m) \leq 3$ in Table \ref{resum_tab3}.  
This correlation function is relevant both for QCD sum-rule analyses
of scalar mesons, a topic of past and present interest \cite{sigma}, and for the
decay of a single-doublet standard-model Higgs boson into a $b\bar{b}$
pair \cite{scalar,higgsier}.  RG-invariance of the correlator ($\mu^2 \mathrm{d}
{\rm Im}\Pi (s) / \mathrm{d}\mu^2 = 0$) implies the following RG-equation for the series
$S [x, L]$ within \tref{scalar_quark_corr} \cite{ces}:
\begin{equation}%6.39
\left[ \frac{\partial}{\partial L} + \beta (x) \frac{\partial}{\partial
x} + 2 \gamma_m (x) \right] S[x, L] = 0,
\tlabel{scalar_quark_RG}
\end{equation}
where $L = \log\left(\mu^2/s\right)$, $x = \alpha_s (\mu) / \pi$, $\beta(x)$ is
the $\beta$-function series \tref{beta_def} and $\gamma_m (x)$ is the 
$\gamma$-function series \tref{mass_RG}.  Substitution of \tref{pert_series} into \tref{scalar_quark_RG} leads
to the following recursion formulae for the elimination of terms
proportional to $x^n L^{n-1}$, $x^n L^{n-2}$, $x^n L^{n-3}$, and $x^n
L^{n-4}$:
\begin{gather}%6.40
0=n T_{n,n} - \beta_0 (n-1) T_{n-1, n-1} - 2\gamma_0 T_{n-1, n-1} 
\tlabel{scalar_qq_RG_e1}
\\
0=(n-1) T_{n, n-1}  -  \beta_0 (n-1) T_{n-1, n-2} - \beta_1 (n-2) T_{n-2, n-
2} -  2\gamma_0 T_{n-1, n-2} - 2\gamma_1 T_{n-2, n-2} 
\tlabel{scalar_qq_RG_e2}
\\
\begin{split}
0=&(n-2) T_{n, n-2}  -  \beta_0 (n-1) T_{n-1, n-3} - \beta_1 (n-2) T_{n-2, n-
3} - \beta_2 (n-3) T_{n-3, n-3}
\\
& -  2\gamma_0 T_{n-1, n-3} - 2\gamma_1 T_{n-2, n-3} - 2\gamma_2 T_{n-
3, n-3} 
\end{split}
\tlabel{scalar_qq_RG_e3}
\\
\begin{split}%6.43
0=&(n-3) T_{n, n-3}  -  \beta_0 (n-1) T_{n-1, n-4} - \beta_1 (n-2) T_{n-2, n-
4} -  \beta_2 (n-3) T_{n-3, n-4} 
\\
&- \beta_3 (n-4) T_{n-4, n-4}
 -  2\gamma_0 T_{n-1, n-4} - 2\gamma_1 T_{n-2, n-4} - 2\gamma_2 T_{n-
3, n-4} - 2\gamma_3 T_{n-4, n-4} 
\end{split}
\tlabel{scalar_qq_RG_e4}
\end{gather}
We follow the usual procedure of
\begin{enumerate}
\item  multiplying \tref{scalar_qq_RG_e1} by $u^{n-1}$ and summing from $n=1$ to
$\infty$,
\item  multiplying \tref{scalar_qq_RG_e2} by $u^{n-2}$ and summing from $n=2$ to
$\infty$,
\item  multiplying \tref{scalar_qq_RG_e3} by $u^{n-3}$ and summing from $n=3$ to
$\infty$, and
\item  multiplying \tref{scalar_qq_RG_e4} by $u^{n-4}$ and summing from $n=4$ to
$\infty$.
\end{enumerate}
Using the definitions \tref{S0_decay_def}, \tref{S1_decay_def}, \tref{S2_decay_def} and \tref{S3_decay_def} 
for $\{S_0, S_1,
S_2, S_3\}$, we then obtain the following four linear differential
equations for these summations:
\begin{gather}%6.44
(1-\beta_0 u) \frac{\mathrm{d} S_0}{\mathrm{d}u} - 2\gamma_0 S_0 = 0, \; \; S_0 [0] = 1
\\
(1-\beta_0 u) \frac{\mathrm{d}S_1}{\mathrm{d}u} - (\beta_0 + 2\gamma_0) S_1 = \beta_1 u
\frac{\mathrm{d}S_0}{\mathrm{d}u} + 2\gamma_1 S_0, \; \; S_1[0]=T_{1,0}
\\
(1-\beta_0 u) \frac{\mathrm{d}S_2}{\mathrm{d}u} - (2\beta_0 + 2\gamma_0) S_2   = \beta_1 u
\frac{\mathrm{d}S_1}{\mathrm{d}u} + \beta_2 u \frac{\mathrm{d}S_0}{\mathrm{d}u}
 +  (\beta_1 + 2\gamma_1) S_1 + 2\gamma_2 S_0, \; \;
S_2[0]=T_{2,0}
\\
\begin{split}
(1-\beta_0 u) \frac{\mathrm{d}S_3}{\mathrm{d}u} - (3\beta_0 + 2\gamma_0) S_3   =&  \beta_1 u
\frac{\mathrm{d}S_2}{\mathrm{d}u} + \beta_2 u \frac{\mathrm{d}S_1}{\mathrm{d}u}
 +  \beta_3 u \frac{\mathrm{d}S_0}{\mathrm{d}u} + (2\beta_1 + 2\gamma_1) S_2
\\
& +  (\beta_2 + 2\gamma_2) S_1 + 2\gamma_3 S_0, \; \; S_3[0]=T_{3,0}
\end{split}
\end{gather}
The solutions to these equations are
\begin{gather}%6.48
S_0 [u] = (1-\beta_0 u)^{-A}
\tlabel{scalar_qq_S0}
\\
S_1 [u] = C_1 (1-\beta_0 u)^{-A} + \frac{[T_{1,0} - C_1 + C_2 \log (1-
\beta_0 u)]}{(1-\beta_0 u)^{A+1}}
\end{gather}
\begin{gather}
\begin{split}
S_2 [u]  = & \frac{D_1}{2} (1-\beta_0 u)^{-A} + \frac{[D_2 - D_3 + D_3 \log (1-
\beta_0 u)]}{(1-\beta_0 u)^{A+1}}
\\
& +  \frac{[T_{2,0} - \frac{D_1}{2} - D_2 + D_3 + D_4 \log (1-\beta_0
u) + \frac{D_5}{2} \log^2 (1-\beta_0 u)]}{(1-\beta_0
u)^{A+2}}
\end{split}
\\
\begin{split}%6.51
S_3 [u]  = & \frac{F_1}{3} (1-\beta_0 u)^{-A} + \frac{[\frac{F_2}{2} -
\frac{F_3}{4} + \frac{F_3}{2} \log (1-\beta_0 u)]}{(1-\beta_0
u)^{A+1}}
\\
& +  \frac{F_4 - F_5 + 2F_6 + (F_5 - 2F_6) \log (1-\beta_0 u) + F_6
\log^2 (1-\beta_0 u)]}{(1-\beta_0 u)^{A+2}}
\\
& + \frac{ \left[ T_{3,0} - \frac{F_1}{3} - \frac{F_2}{2} + \frac{F_3}{4} - F_4 + F_5 -
2F_6 + F_7 \log (1-\beta_0 u)
 +   \frac{F_8}{2} \log^2 (1-\beta_0 u) + \frac{F_9}{ 3} \log^3 (1-\beta_0
u) \right]}{  (1-\beta_0 u)^{A+3}},
\end{split}
\tlabel{scalar_qq_S3}
\end{gather}
where
\begin{gather}%6.52
A = 2\gamma_0  \beta_0
\\
C_1 = 2\beta_1 \gamma_0 / \beta_0^2 - 2\gamma_1 / \beta_0
\\
C_2 = -2\beta_1 \gamma_0 / \beta_0^2
\\
D_1 = \left[ C_1 \left(A\beta_1 - \beta_1 - 2\gamma_1\right) + A\beta_2 - 2\gamma_2
\right] / \beta_0
\\
D_2 = \left[ C_1 \left(2\gamma_1 - 2A\beta_1\right) - C_2 \beta_1 - A\beta_2 +
T_{1,0} \left(A\beta_1 - 2\gamma_1\right) \right] / \beta_0
\\
D_3 = C_2 \left(A\beta_1 - 2\gamma_1\right) / \beta_0
\\
D_4 = \beta_1 \left[ C_2 - (A+1)\left(T_{1,0} - C_1\right) \right] / \beta_0
\\
D_5 = -C_2 \beta_1 (A+1) / \beta_0
\\
F_1 = \left[ C_1 \left(A\beta_2 - 2\gamma_2 - \beta_2\right) + D_1 \left(A\beta_1 / 2 -
\gamma_1 - \beta_1\right) + A\beta_3 - 2\gamma_3 \right] / \beta_0
\\
\begin{split}
F_2  =&  \frac{1}{ \beta_0} \biggl[ D_3 \left(2\gamma_1 - A\beta_1\right) + D_2 \left(A\beta_1 - \beta_1 -
2\gamma_1\right) - \frac{D_1}{2} A\beta_1  - C_2 \beta_2 \biggr.
\\
 &\qquad \biggl.+   C_1\left(2\gamma_2 - 2A\beta_2\right) + T_{1,0}\left(A\beta_2 - 2\gamma_2\right) -
A\beta_3 \biggr]
\end{split}
\\
F_3 = \left[D_3 \left(A\beta_1 - \beta_1 - 2\gamma_1\right) + C_2 \left(A\beta_2 -
2\gamma_2\right)\right] / \beta_0
\\
\begin{split}
F_4  =&  \frac{1}{ \beta_0} \biggl[ \left(T_{2,0} - D_1 / 2 - D_2 + D_3\right)\left(A\beta_1 - 2\gamma_1\right) -
D_4 \beta_1 
 +  D_3 (A+2)\beta_1 - D_2 (A+1) \beta_1 \biggr.
\\
&\qquad \biggl.+ C_2 \beta_2
 +   C_1 (A+1) \beta_2 - T_{1,0} (A+1) \beta_2 \biggr]
\end{split}
\\
F_5 = \left[ D_4 \left(A\beta_1 - 2\gamma_1\right) - D_3 (A+1) \beta_1 - D_5 
\beta_1 - C_2 (A+1) \beta_2 \right] / \beta_0
\\
F_6 = D_5 \left(A\beta_1 / 2 - \gamma_1\right) / \beta_0
\\
F_7 = \beta_1 \left[ D_4 - (A+2)\left(T_{2,0} - D_1 / 2 - D_2 + D_3\right)\right] /
\beta_0
\\
F_8 = \beta_1 \left[ D_5 - D_4(A+2)\right] / \beta_0
\\
F_9 = -\beta_1 D_5 (A+2) / \left(2\beta_0\right)
\end{gather}

The ${\cal{O}}(N^3 L)$ RG-summation of the series $S[x(\mu), \;
\log(\mu^2 / s)]$ appearing in the correlator \tref{scalar_quark_corr} is then found to be
\begin{equation}%6.69
\begin{split}
S_{RG\Sigma}^{N^3 L}  =&  S_0 \left[ x(\mu)\log\left(\frac{\mu^2}{ s}\right) \right] +
x(\mu) S_1 \left[ x(\mu) \log\left(\frac{\mu^2}{ s}\right) \right]
\\
& +  x^2 (\mu) S_2 \left[ x(\mu) \log\left(\frac{\mu^2}{ s}\right) \right] + x^3 (\mu)
S_3 \left[ x(\mu) \log\left(\frac{\mu^2}{ s}\right) \right]
\end{split}
\end{equation}
where the RG-summations $S_0, S_1, S_2$ and $S_3$ are given by \tref{scalar_qq_S0}--\tref{scalar_qq_S3}
with $u = x(\mu) \log (\mu^2 / s)$.

\begin{figure}[hbt]
\centering
\includegraphics[scale=0.7]{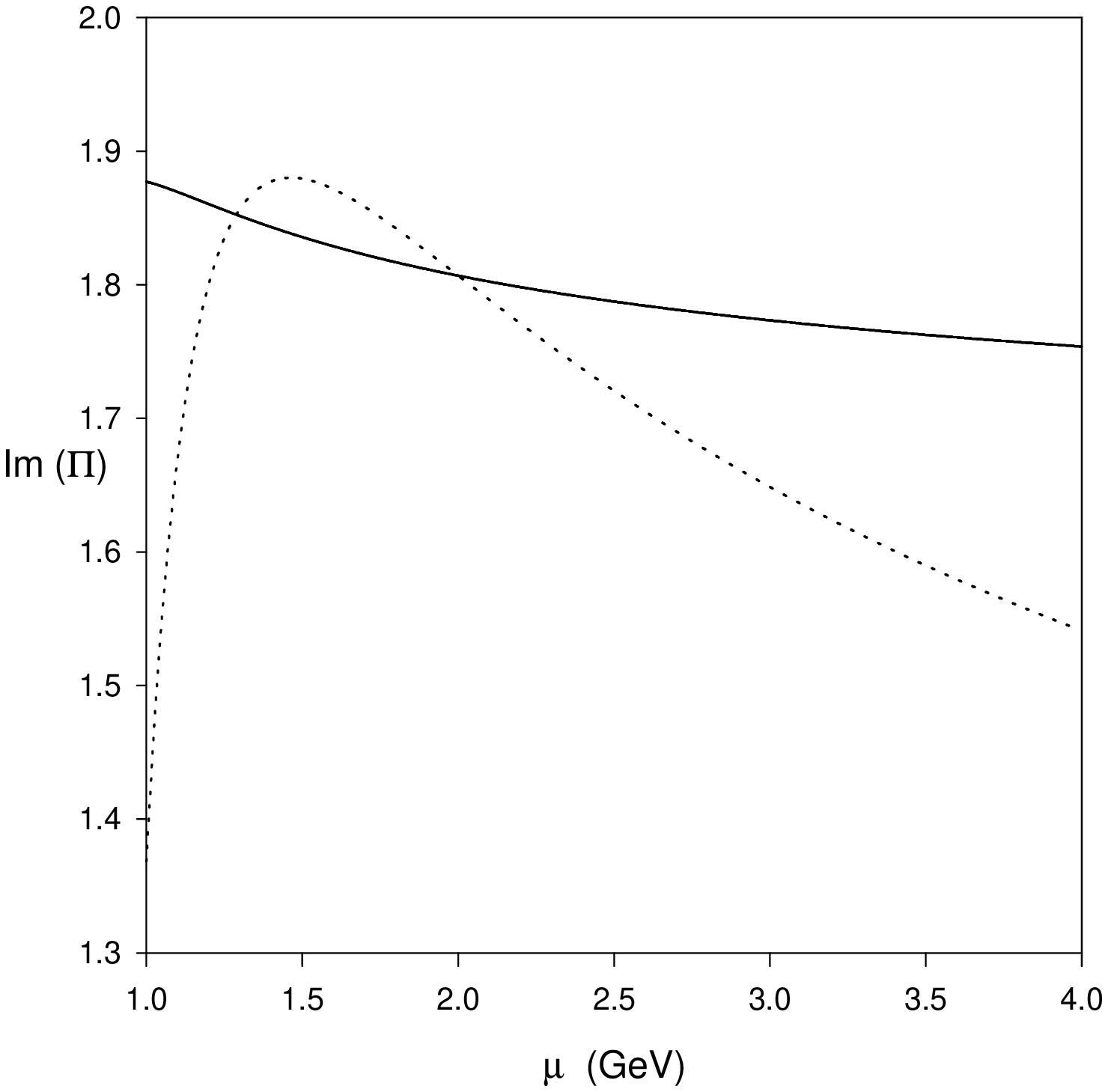}
\caption{
 The imaginary part of the fermionic scalar-current correlation function (\protect\ref{scalar_quark_corr}), as
          obtained from $m^2(\mu)\,S^{NNL}$ (dotted line) and from  $m^2(\mu)\,S^{NNL}_{RG\Sigma}$  (solid line) 
with $n_f = 3$
          and  $\sqrt{s} = 2\,{\rm  GeV}$.
}
\label{resum_f16}
\end{figure}

In Figure \ref{resum_f16} we compare the $\mu$-dependence of the RG-summed scalar
fermionic-current correlator,
\begin{equation}%6.70
{\rm Im}\Pi_{RG\Sigma}^{N^3 L} \sim m^2 (\mu) S_{RG\Sigma}^{N^3 L}
\label{Imqq_sum}
\end{equation}
to that of the correlator \tref{scalar_quark_corr} when truncated after its fully-known
${\cal{O}}(x^3)$ contributions,
\begin{equation}%6.71
{\rm Im}\Pi^{N^3 L} \sim m^2 (\mu) \sum_{n=0}^3 \sum_{m=0}^n T_{n, m} x^n
(\mu) \left[ \log\left(\frac{\mu^2}{s}\right)\right]^m,
\label{Imqq_trunc}
\end{equation}
over the range $\sqrt{s}/2 \leq \mu \leq 2\sqrt{s}$ with $\sqrt{s} = 2\;
\,{\rm GeV}$.  The coefficients $T_{n,m}$ appearing in (6.71) are given in Table \ref{resum_tab3}.  
We choose to work in the $\sqrt{s} = 2\,{\rm GeV} \; \; n_f=3$ regime
appropriate for QCD sum rule applications, where the couplant $x(\mu)$
is large.  The evolution of $x(\mu)$ is assumed to proceed via the $n_f
= 3$ four-loop $\beta$-function with initial condition $x(m_\tau) =
0.33/\pi$ \cite{alpha}.  The running mass $m(\mu)$ is normalized to $1\;
\,{\rm GeV}$  at $\mu = m_\tau$  to facilitate comparison of 
\tref{Imqq_sum} to \tref{Imqq_trunc}.  In Figure \ref{resum_f16}, the unsummed
correlator is seen to achieve a sharp maximum near $1.5 \,{\rm GeV}$,
followed by a precipitous fall as $\mu$ approaches $1\,{\rm GeV}$ from above.
By contrast, the RG-summed correlator exhibits a much flatter profile,
falling from $1.88 \,{\rm GeV}^2$ to $1.75 \,{\rm GeV}^2$ as $\mu$ increases from
$1$ to $4 \,{\rm GeV}$.  As in Section \ref{scalar_glue_sec} for the gluonic scalar-current
correlator, these results indicate that even $N^3 L$-order expressions for the
perturbative contribution to QCD correlation functions exhibit
substantial $\mu$-dependence. Such dependence, which we have shown to be largely 
eliminated via the RG-summation process, would otherwise percolate through QCD 
sum-rule integrals as spurious sensitivity on the theory side to the Borel parameter.  
The incorporation of RG-summed correlators within QCD sum-rule
extractions of lowest-lying scalar resonances is currently under
investigation.

\section{Summary}
In this paper we have explicitly summed all RG-accessible logarithms
within a number of perturbative processes known to at least two
nonleading orders, a procedure originally advocated by Maxwell
\cite{maxwell}.  As anticipated, we have found the dependence on
the renormalization scale $\mu$ in every case examined to be considerably diminished over that
of the original series' known terms.

In semileptonic $b \rightarrow u$ decays in the fully $\overline{\rm{MS}}$
scheme (Section \ref{bu_MS_sec}), we observe the intriguing possibility that PMS/FAC
criteria for the unsummed series truncated to a given order may
anticipate the RG-summed series for the next order of perturbation
theory.\footnote{Although this result incorporates a Pad\'e estimate
for an RG-inaccessible ${\cal{O}}(x^3)$ coefficient, this estimate occurs in
both the unsummed and RG-summed expression.} This behaviour, however,
is not evident in the other processes we consider.  For the vector
correlation function (Section \ref{vec_corr_sec}), PMS/FAC criteria for the unsummed
series truncated to a given order coincide closely with the RG-summed
series for that same order, but do not anticipate the level of the 
next-order RG-summation.  In the pole-mass scheme version of semi-leptonic 
B-decays (Section \ref{bu_pole_sec}), PMS and FAC criteria do not appear applicable
to the unsummed series, which monotonically increase with the
renormalization scale $\mu$.  Indeed, one of the virtues of RG-summation
is the sensible scale-independent results it provides for inclusive
semileptonic B-decays in the pole mass scheme, a scheme whose unsummed
expressions for $b \rightarrow u$ are already known to be problematical
\cite{vanritbergen}.

The $\mu$-independence of RG-summation, particularly for the 
vector-current correlation function (Section \ref{vec_corr_sec}), is seen to justify the
prescription of zeroing all logarithms by setting the renormalization
scale to $\mu^2$ equal to the kinematic variable $s$.  This prescription
necessarily equates the unsummed and RG-summed series, and, since the
RG-summed series is virtually independent of $\mu$, the zeroing of
logarithms in the unsummed series equilibrates it to the flat RG-summation 
level we obtain.

In Section \ref{pert_sec}, RG-summation is also applied to the perturbative
contributions to the momentum-space QCD static potential, the decay-rate
of a standard-model Higgs to two gluons, the Higgs-mediated 
cross-section $\sigma(WW \rightarrow ZZ)$, and to two scalar-current
correlation functions.  Examination of these last two quantities in the
low-$s$ region appropriate for QCD sum-rules suggests the utility of 
RG-summation is reducing the unphysical scale-dependence of the
perturbative QCD contributions to the field-theoretical side of 
sum-rules in these channels.

\section{Acknowledgements}
We are grateful for research support from Leadership Mt.\ Allison, the
International Collaboration Programme 2001 of Enterprise Ireland,
2001 Faculty Summer Grant from SUNY Institute of Technology,
and
the International Opportunity Fund of the Natural Sciences and
Engineering Research Council of Canada.  We also acknowledge the
hospitality of the High Energy Theory Group at KEK, Tsukuba, Japan,
where this research was initiated.  Finally, we are grateful to Audrey
Kager for valuable contributions to the preparation of this research in
manuscript form.

\section*{Appendix: An Alternative Closed-Form Summation Procedure}
\renewcommand{\theequation}
{A.\arabic{equation}}
\setcounter{equation}{0}
The body of our paper has addressed the evaluation of 
$S_n\left[xL\right]=\sum_{k=0}^\infty T_{n+k,k}\left(xL\right)^k$,
where the full perturbative series 
$S(x,L)=\sum_{n=0}^\infty S_n\left[xL\right]x^n$.  It is, however, also 
possible to group the terms within $S(x,L)$, as defined by eq.\ \tref{pert_series}, such that
the dependence of each series term on $x$ and $L$ fully factorises:
\begin{gather}
S(x,L)=\sum_{n=0}^\infty R_n(x) L^n\quad ,
\tlabel{fac_form}
\\
R_n(x)=\sum_{k=n}^\infty T_{k,n} x^k\quad .
\end{gather}
Let us suppose, for example, that the series $S(x,L)$ satisfies the RGE \tref{decay_RGE}
appropriate to $b$-decays. By substituting \tref{fac_form} into \tref{decay_RGE}, we find that
\begin{equation}
R_{n+1}(x)=\frac{1}{n+1}\left[\frac{1}{2\gamma(x)-1}\right]
\left[\beta(x)\frac{\mathrm{d}}{\mathrm{d}x}+5\gamma(x)\right]R_n(x) \quad ,
\tlabel{fac_RG}
\end{equation}
where we have relabeled the anomalous mass dimension $\gamma_m(x)\to\gamma(x)$ to avoid any misinterpretation of the label $m$ as a subscript.  If
\begin{equation}
R_n(x)\equiv \exp{\left(
-\int\limits^x\frac{5\gamma\left(x'\right)}{\beta\left(x'\right)}\,\mathrm{d}x'\right)}P_n(x)
\quad ,
\tlabel{Rn_def}
\end{equation}
then the recursion relation \tref{fac_RG} implies that
\begin{equation}
P_{n+1}(x)=\frac{1}{n+1}\left(\frac{\beta(x)}{2\gamma(x)-1}\right)\frac{\mathrm{d}}{\mathrm{d}x}P_n(x)
\quad .
\tlabel{fac_P_rec}
\end{equation}
If one defines $x$ to be implicitly a function of $y$ via the equation
\begin{equation}
\frac{\mathrm{d}}{\mathrm{d}y}x(y)=\frac{\beta(x)}{2\gamma(x)-1}\quad ,
\tlabel{y_def}
\end{equation}
then \tref{fac_P_rec} simplifies to 
\begin{equation}
P_{n+1}\left(x(y)\right)=\frac{1}{n+1}\frac{\mathrm{d}}{\mathrm{d}y}P_n\left(x(y)\right)
\quad ,
\end{equation}
in which case
\begin{equation}
\sum_{n=0}^\infty P_n\left(x(y)\right)L^n=\sum_{n=0}^\infty
\left(
\frac{1}{n!}L^n\frac{\mathrm{d}^n}{\mathrm{d}y^n}\right)P_0\left(x(y)\right)
=P_0\left(x(y+L)\right)\quad .
\end{equation}
This last result implies via \tref{Rn_def} that the series $S(x,L)$ is fully determined by knowledge of 
the log-free summation $R_0$, {\it i.e.} that
\begin{equation}
S(x,L)=\exp{\left[\int\limits_{x(y)}^{x(y+L)}\frac{5\gamma\left(x'\right)}{\beta\left(x'\right)}\,\mathrm{d}x' 
\right]}
R_0\left(x(y+L)\right)
\tlabel{log_free}
\end{equation}
where $x(y)$ is defined implicitly by the constraint
\begin{equation}
y=\int\limits^x\frac{2\gamma\left(x'\right)-1}{\beta\left(x'\right)}\,\mathrm{d}x'
\tlabel{y_eqn}
\end{equation}
obtained by integrating \tref{y_def}.

Using lowest-order expressions $\beta(x)=-\beta_0x^2$ and $\gamma(x)=-\gamma_0x$, we find from 
\tref{y_eqn} that
\begin{equation}
y=\frac{1}{\beta_0}\left(-\frac{1}{x}+2\gamma_0\log(x)\right)+K\quad .
\tlabel{y_res}
\end{equation}
If we set 
\begin{equation}
x=\frac{1}{2\gamma_0W}\quad ,\quad K=\frac{2\gamma_0}{\beta_0}\log{\left(2\gamma_0\right)}
\end{equation}
we find from \tref{y_res} that
\begin{equation}
W\mathrm{e}^W=\exp{\left(-\frac{\beta_0 y}{2\gamma_0}\right)}\quad .
\tlabel{W_fn}
\end{equation}
Eq.\ \tref{W_fn} is the defining relationship for the Lambert $W$-function
$W\left[\exp{\left(-\beta_0 y/2\gamma_0\right)} \right]$, as discussed in ref.\ \cite{corless}. 
Since
\begin{equation}
x(y)=\frac{1}{2\gamma_0W\left[\exp{\left(-\beta_0 y/2\gamma_0\right)}\right]}
\end{equation}
in the approximation $\beta(x)\cong-\beta_0 x^2$, $\gamma(x)\cong-\gamma_0x$, we then find from
\tref{log_free} that
\begin{equation}
S(x,L)=\left[
\frac{W\left[\exp{\left(-\beta_0 y/2\gamma_0\right)}\right]}{W\left[\exp{\left(-\beta_0 (y+L)/2\gamma_0\right)} \right]}
\right]^{5\gamma_0/\beta_0}
R_0\left(\frac{1}{2\gamma_0W\left[\exp{\left(-\beta_0 (y+L)/2\gamma_0\right)} \right]}\right)
\quad ,
\end{equation}
where 
\begin{equation}
R_0(x)=\sum_{k=0}^\infty T_{k,0}x^k=S(x,0)
\tlabel{R0_def}
\end{equation}
and where 
\begin{equation}
y=-\frac{1}{\beta_0 x}+\frac{2\gamma_0}{\beta_0}\log\left(2\gamma_0x\right)
\quad .
\end{equation}

For the RGE \tref{bc_RGE}, corresponding to \tref{decay_RGE} with $\gamma_m(x)$ chosen to be zero, the solution 
\tref{log_free} still applies provided  $x(y)$ is defined implicitly via the constraint
\begin{equation}
y=-\int\limits^x\frac{\mathrm{d}x'}{\beta\left(x'\right)}\quad .
\end{equation}
In the approximation $\beta(x)=-\beta_0x^2$, one can choose
\begin{equation}
x(y)\equiv -\frac{1}{\beta_0y}
\end{equation} 
in which case
\begin{equation}
S(x,L)=R_0\left(\frac{x}{1-\beta_0Lx}\right)\quad ,
\end{equation}
where the function $R_0$ is as defined via \tref{R0_def}.

\newpage

\begin{table}%1
\centering
\begin{tabular}{c|ccc}
& $n_f = 3$ & $n_f=4$ & $n_f = 5$\\
\hline
$\beta_0$ & 9/4 & 25/12 & 23/12\\
$\beta_1$ & 4 & 77/24 & 29/12\\
$\beta_2$ & 3863/384 & 21943/3456 & 9769/3456\\
$T_{2,0}$ & 1.63982 & 1.52453 & 1.40924\\
$T_{2,1}$ & 9/4 & 25/12 & 23/12\\
$T_{3,0}$ & -10.2839 & -11.6856 & -12.8046\\
$T_{3,1}$ & 11.3792 & 9.56054 & 7.81875\\
$T_{3,2}$ & 81/16 & 625/144 & 529/144\\
$G$ & -16/9 & -77/50 & -29/23\\
$P$ & $-\frac{3397}{2592}$ & $-\frac{121687}{180000}$ & $\frac{17521}{152352}$\\
$U$ & -8.99096 & -7.06715 & -5.14353\\
$V$ & 512/81 & 5929/1250 & 1682/529
\end{tabular}
\caption{Constants for determining the ${\cal{O}}(N^3 L)$ RG-summation
of the $N^3 L$-order vector-current correlation function.}
\label{resum_tab1}
\end{table}

\begin{table}%2
\centering
\begin{tabular}{c|ccccc}
& $n_f=2$ & $n_f = 3$ & $n_f = 4$ & $n_f = 5$ & $n_f = 6$\\
\hline\\
$T_{1,0}$ & $\frac{6919}{348}$ & $\frac{659}{36}$  & $\frac{4999}{300}$ & $\frac{4123}{276}$ & $\frac{367}{28}$\\
\\
$T_{1,1}$ & $\frac{29}{6}$ & $\frac{9}{2}$  & $\frac{25}{6}$ & $\frac{23}{6}$ & $\frac{7}{2}$\\
\\
$T_{2,0}$ & 246.434 & 197.515  & 150.210 & 104.499 & 60.3685\\
\\
$T_{2,1}$ & $\frac{7379}{48}$ & $\frac{2105}{16}$ & $\frac{1769}{16}$ & $\frac{4355}{48}$ & $\frac{1153}{16}$\\
\\
$T_{2,2}$ & $\frac{841}{48}$ & $\frac{243}{16}$  & $\frac{625}{48}$ & $\frac{529}{48}$ & $\frac{147}{16}$ 
\end{tabular}
\caption{${\rm NNL}$-order series coefficients within the gluonic scalar current correlation function.}
\label{resum_tab2}
\end{table}

\begin{table}%3
\centering
\begin{tabular}{c|ccc}
& $n_f = 3$ & $n_f=4$ & $n_f = 5$\\
\hline
$T_{0,0}$  & 1       &  1         & 1 \\
$T_{1,0}$  & 17/3    &  17/3      & 17/3\\
$T_{1,1}$  & 2       &  2         & 2\\
$T_{2,0}$  & 31.8640 &  30.5054   & 29.1467\\
$T_{2,1}$  & 95/3    &  274/9     & 263/9\\
$T_{2,2}$  & 17/4    &  49/12     & 47/12\\
$T_{3,0}$  & 89.1564 &  65.1980   & 41.7576\\
$T_{3,1}$  & 297.596 &  267.589   & 238.381\\
$T_{3,2}$  & 229/2   &  22547/216 & 10225/108\\
$T_{3,3}$  & 221/24  &  1813/216  & 1645/216\\
$\beta_3$  & 47.2280 &  31.3874   & 18.8522\\
$\gamma_3$ & 44.2628 &  27.3028   & 11.0343
\end{tabular}
\caption{$N^3 L$-order series coefficients within the fermionic scalar current correlation function,
as calculated in \cite{scalar}.  Also listed are the four-loop $\beta$-function \cite{beta}
and $\gamma$-function \cite{gamma} coefficients $\beta_3$ and $\gamma_3$ required for
the evaluation of the series $S_3$.}
\label{resum_tab3}
\end{table}

\clearpage

\clearpage

\clearpage

\clearpage

\clearpage

\clearpage

\clearpage

\clearpage

\clearpage

\clearpage

\clearpage

\clearpage

\clearpage

\clearpage

\clearpage

\end{document}